\documentclass[aps,pra]{revtex4}

\usepackage{amsmath}
\usepackage{graphicx}
\usepackage{epsfig}
\usepackage{bm}

\newcommand{\be}{\begin{equation}}
\newcommand{\ee}{\end{equation}}

\newcommand{\beq}{\begin{eqnarray}}
\newcommand{\eeq}{\end{eqnarray}}

\def\H1{\widehat{H}_1}

\begin{document}

\title{Exact methods in analysis of nonequilibrium dynamics of integrable models:
application to the study of correlation functions in nonequilibrium 1D
Bose gas}
\author{Vladimir Gritsev$^{1,2}$}
\author{Timofei Rostunov$^{1}$}
\author{Eugene Demler$^1$}
\affiliation{$^1$Lyman Laboratory of Physics, Physics Department,
Harvard University, 17 Oxford Street, Cambridge MA, 02138, USA\\
$^2$Physics Department, University of Fribourg, Chemin du Musee 3,
1700 Fribourg, Switzerland}
\begin{abstract}{
In this paper we study nonequilibrium dynamics of one dimensional
Bose gas from the general perspective of dynamics of integrable
systems. After outlining and critically reviewing methods based on
inverse scattering transform, intertwining operators, q-deformed
objects, and extended dynamical conformal symmetry,  we focus on the
form-factor based approach. Motivated by possible applications in
nonlinear quantum optics and experiments with ultracold atoms, we
concentrate on the regime of strong repulsive interactions. We
consider dynamical evolution starting from two initial states:  a
condensate of particles in a state with zero momentum and a
condensate of particles in a gaussian wavepacket in real space.
Combining the form-factor approach with the method of intertwining
operator we develop a numerical procedure which allows explicit
summation over intermediate states and analysis of the time
evolution of non-local density-density correlation functions. In
both cases we observe a tendency toward formation of crystal-like
correlations at intermediate time scales.

}
\end{abstract}
\maketitle

\section{Introduction}
Quantum many-body systems in low dimensions can not typically be
described using mean-field approaches. This makes analysis of their
non-equilibrium dynamics particularly challenging. However following
demand from recent experiments, certain progress has been achieved
in developing theoretical methods which can address this problem. In
this paper we use Bethe ansatz solution to study nonequilibrium
dynamics of the one dimensional Bose gas interacting via contact
interaction. This microscopic model of $\delta$-interacting bosons
is called the Lieb-Liniger (LL) model
\cite{LL},\cite{McG},\cite{Ber}. It belongs to a general class of
models which can be studied using Bethe ansatz solution \cite{BIK}.
These exactly solvable models are characterized by an infinite
number (in the thermodynamic limit) of conserved quantities which
originate from the nature of collision processes. In the LL model
the two particle collisions can not change momenta of scattering
particles but only give rise to a phase shift. Moreover any
many-body collision is factorizable into a sequence of two-body
scattering events. While implications of the exact solution for
thermodynamic properties have been discussed
before\cite{LL,BIK,gaudin}, their consequences for non-equilibrium
dynamics are not known. This will be the central question of our
paper.

Systems of one dimensional Bose gases with contact interactions have
been recently realized using ultracold atoms
\cite{bloch,KWW1,KWW2,schmiedmayer,amsterdam} and current
experiments allow wide control over parameters of the microscopic
Hamiltonian (see \cite{BDZ},\cite{Lew} for recent review). For
example, the effective strength of the repulsive interaction can be
tuned either by changing the density of the atomic cloud, or by
modifying the strength of the transverse confinement, or by changing
the scattering length using magnetically tuned Feshbach resonance.
In particular a very interesting regime can be achieved when
repulsion is so strong that bosonic atoms become essentially
impenetrable\cite{TG} and the system undergoes effective
fermionization\cite{KWW1,KWW2,bloch}. Recent studies of
thermodynamic properties demonstrated excellent agreement between
experiments and the exact solution \cite{KWW2,amsterdam}. However
parameters of the Hamiltonian can also be changed dynamically. In
this case one needs to study dynamics starting with the initial
state which is not an eigenstate of the Hamiltonian. Generally such
dynamics involves superposition of coherent evolution of all
eigenstates of the system. This is the problem that we address in
this paper.

Another class of systems  where the LL model appears naturally is
light propagation in one dimensional photonic fibers. It has been
known since the sixties (see e.g. ref \cite{Zaharov-Shabat}) that
propagation of classical pulses of electromagnetic waves in one
dimensional nonlinear medium can be described by the nonlinear
Schroedinger equation, which can be interpreted as the operator
equation of motion arising from the LL microscopic Hamiltonian.   In
this case nonlinearity is proportional to the nonlinear
polarizability of the medium, $\chi^{(3)}$ (see e.g. Ref
\cite{boyd92}). Typical propagation problem in one dimensional
nonlinear fiber corresponds to the classical limit of the quantum LL
model subject to nonequilibrium boundary and/or initial conditions
\cite{hafezi}. Earlier analysis assumed that photon nonlinearities
come from interaction of photons with two level atoms, in which case
nonlinearity corresponds to the attractive interaction in the LL
model \cite{LH}. Corresponding quantum models have bound states
which eventually lead to formation of solitons and classical regime
\cite{WS}. Earlier analysis of optical systems also assumed regime
of weak nonlinearities since increasing nonlinearities in two level
systems would lead to strong photon losses. However recent
techniques utilizing electromagnetically induced transparency
\cite{EIT} make it possible not only to realize effective repulsion
between photons but also to increase dramatically the strength of
nonlinearities\cite{Chang}. Optical realizations  of the LL model
should allow direct measurements of both the first and the second
order coherences \cite{Glauber},
$g^{(1)}(t,\tau)=\langle\Psi^{\dag}(t)\Psi(t+\tau)\rangle$ and
$g^{(2)}(t,\tau)=\langle\rho(t)\rho(t+\tau)\rangle$, where $\Psi(t)$
is the amplitude of the propagating electromagnetic field and
$\rho(t)=\Psi^{\dag}(t)\Psi(t)$ is the field intensity, measuring
the number of photons.

There are three main reasons why we undertake a detailed analysis of
the non-equilibrium quantum dynamics of the LL model. The {\it
first} reason is that dynamics of integrable models, such as the LL
model, should exhibit special features originating from an infinite
number of integrals of motion. One manifestation of constraints on
the phase space available for dynamics should be the absence of
thermalization and ergodicity \cite{GGE,schmiedmayer,BS}.
Conservation laws originating from integrability also lead to the
so-called Mazur inequalities \cite{Mazur}, which should result in
special transport properties. For example, the possibility of an
infinite Drude weight has been discussed for several integrable
models of lattice fermions \cite{Zotos}. Our {\it second} motivation
for analysis of the LL model is to use it as a testing ground for
developing new theoretical methods and techniques, which can then be
applied to a broader class of models and systems. Our work should
provide a general framework for understanding dynamics of integrable
systems and for future developments of non-perturbative methods in
the study of non-equilibrium dynamics.
We emphasize that our analysis  uses special properties of exactly
solvable models, and the methods we apply here are fundamentally
different from the more conventional perturbative approaches
discussed in Refs. \cite{Keldysh},\cite{Schoeller}. Finally we point
out that recent progress in the areas of ultracold atoms, quantum
optics, and low-dimensional strongly-correlated materials makes it
possible to fabricate concrete physical systems, which can be
accurately described by the LL model. Thus our {\it third} reason
for analyzing this model is that theoretical predictions for the
time-dependent evolution of correlation functions can be measured
directly in experiments. Specific system in which one can realize
dynamical experiments discussed in our paper are one dimensional
Bose gases in optical lattices and magnetic microtraps (see Refs.
\cite{tubingen,schmiedmayer2} for a review). Realization of such
experiments should provide a unique opportunity to study
non-equilibrium dynamics of strongly correlated exactly solvable
systems.

 In this paper we calculate  time evolution of the
density-density correlation function for  two different
condensate-like initial states: when all particles are in a state
with zero momentum and when all particles are in a Gaussian
 wave packet in real space.  There are two main reasons why we focus
on these initial states. Firstly, both of these states correspond to important
physical systems. A condensate of particles in the zero-momentum state
provides a natural initial state for the discussion of dynamics of
a Bose condensate when interaction is suddenly switched on.
The Gaussian wave packet in
real space is a prototypical example of the photonic pulse in
quantum optics. Secondly, considerable progress can be made in the
analytic calculations of the time
evolution arising from these two states. Also numeical calculations can be
done quite efficiently. We emphasize here
that it is {\it not} the case for a generic initial
state. Understanding dynamics of the generic initial
state requires developing additional theoretical tools.

This paper is organized as follows. Section II provides a brief {\it
critical} overview of several different methods which can be used to
describe non-equilibrium dynamics of integrable quantum systems such
as the LL model. Section III provides an in-depth discussion of one
of these approaches, the so-called  form-factor technique, which
relies on numerical summation over intermediate states. The overlap
with the initial states in the strongly interacting regime is found
using the method of intertwining operator. In section IV we focus on
the non-equilibrium dynamics of the many-body system described by
the LL model. There we choose two particular initial states as
explained above. Summary of our results and conclusions are given in
Section V.  To provide additional background to the readers, in
appendices we provide basic facts about the algebraic Bethe Ansatz
formalism and the inverse scattering transform for the model of one
dimensional Bose gas with both repulsive and attractive
interactions.


\section{A review of methods to investigate nonequilibrium dynamics of integrable models}

In this paper we analyze the time evolution of correlation functions
in integrable models when the  initial state is not an eigenstate of
the Hamiltonian. Here we give an overview of (some of) the possible
approaches to studying nonequilibrium dynamics of integrable
systems. We only discuss approaches which explicitely use the
property of integrability and exclude more conventional techniques
which rely on approximations and perturbative expansions, such as
the Keldysh formalism \cite{Keldysh} . Although the focus of this
paper is on the 1D Bose gas, methods discussed here can be applied
to other exactly-solvable models, e.g. for the Gaudin-type
integrable models \cite{FCC}.

The Hamiltonian of the 1D interacting Bose gas in a finite size
system has the following form \cite{BIK}
\beq\label{Hc}
H_{c}=\int_{0}^{L} dx
[\partial_{x}\Psi^{\dag}(x)\partial_{x}\Psi(x)+c\Psi^{\dag}(x)\Psi^{\dag}(x)\Psi(x)\Psi(x)]
\eeq
Here $\Psi(x)$ is a bosonic field in the second quantized notations,
$c$ is an interaction constant, $L$ is the size of the system. The
equation of motion for this Hamiltonian is the Nonlinear Schrodinger (NS)
equation.
The first quantized version of this Hamiltonian  corresponds
to the Lieb-Liniger model of Bose gas interacting with the contact
interaction. The Bethe ansatz solution of this model has been given in refs.
\cite{LL},\cite{McG},\cite{Ber},\cite{gaudin}.

Below we discuss  several possible approaches to
the study of nonequilibrium dynamics of nonlinear integrable models, such as the
LL model (\ref{Hc}).

\subsection{Quantum inverse scattering method\label{sec:inv-scatt}}

One approach to analyzing dynamics of model (\ref{Hc}) would be to
use the formalism of the inverse scattering problem. This method
relies on the solution of the quantum version of the
Gelfand-Levitan-Marchenko equation (see Appendix A). From the
Inverse Scattering Transform (see e.g. \cite{Thacker} for review) it
is known that the solution is given by the following infinite series
representation
\beq\label{glmexp}
\Psi(x)=\sum_{N=0}^{\infty}\int\prod_{i=1}^{N}\frac{dp_{i}}{2\pi}\prod_{j=0}^{N}\frac{dk_{j}}{2\pi}g_{N}(\{p\},\{k\};x)R^{\dag}(p_{1})\cdots
R^{\dag}(p_{N})R(k_{N})\cdots R(k_{0})
\eeq
where the operators $R(p)$ diagonalize the problem on the infinite
interval $[-\infty,\infty]$. The function $g_{N}$ can be
cast into different forms \cite{CTW}. One specific representation is given by
\beq\label{gN}
g_{N}(\{p\},\{k\};x)=\frac{(-c)^{N}\exp[ix(\sum_{0}^{N}k_{i}-\sum_{1}^{N}p_{i})]}{\prod_{m=1}^{N}(p_{m}-k_{m}-i\epsilon)(p_{m}-k_{m-1}-i\epsilon)}
\eeq
The above perturbative expansion is a quantum version of Rosales
expansion \cite{Rosales},\cite{davies}. Thus,
\beq
\Psi(x)=\int\frac{d\xi_{1}}{2\pi}R(\xi_{1})e^{i\xi_{1}x}+c^{2}\int
\frac{d\xi_{1}}{2\pi}\int \frac{d\xi_{2}}{2\pi}\int
\frac{d\xi_{3}}{2\pi}\frac{R^{\dag}(\xi_{2})R(\xi_{1})R(\xi_{3})e^{i(\xi_{1}-\xi_{2}+\xi_{3})}}{(\xi_{2}-\xi_{1}-i\epsilon)(\xi_{3}-\xi_{2}+i\epsilon)}+\cdots
\eeq

The inverse expression (direct Gelfand-Levitan transform) is also
available (see \cite{Sklyanin1} for the finite interval),
\beq
R^{\dag}(\xi)=\frac{1}{\sqrt{2\pi}}\int dx
\Psi^{\dag}(x)e^{-iqx}+\frac{1}{\sqrt{2\pi}}\int
dx_{1}dx_{2}dx_{3}g_{2}(x_{1},x_{2},x_{3};q)\Psi^{\dag}(x_{1})\Psi^{\dag}(x_{2})\Psi(x_{3})+...
\eeq
where e.g. \cite{HWW}
$g_{2}(x_{1},x_{2},x_{3};q)=c\theta(x_{2}-x_{3})\theta(x_{3}-x_{1})\exp(-iq(x_{1}+x_{2}-x_{3}))$,
$g_{4}(x_{1},x_{2},x_{3},x_{4},x_{5};q)=c^{2}\theta(x_{3}-x_{5})\theta(x_{5}-x_{2})\theta(x_{2}-x_{4})\theta(x_{4}-x_{1})\exp(-iq(x_{1}+x_{2}+x_{3}-x_{4}-x_{5}))$.
Here the operators of reflection coefficient  $R(\xi),
R^{\dag}(\xi)$ diagonalize the Hamiltonian and satisfy the
Zamolodchikov-Faddeev algebra (see Appendix A for more details)
\beq\label{Rbasis}
~[H,R^{\dag}(\xi)]&=&\xi^{2}R^{\dag}(\xi),\\
R(\xi)R(\xi')&=&S(\xi'-\xi)R(\xi')R(\xi),\qquad
R(\xi)R^{\dag}(\xi')=S(\xi-\xi')R^{\dag}(\xi')R(\xi)+2\pi
\delta(\xi-\xi').
\eeq
where the scattering matrix is
$S(\xi-\xi')=(\xi-\xi'-ic)/(\xi-\xi'+ic)$. These
relations make the problem of finding time evolution easy: the factor
$\exp[ix(\sum_{0}^{N}k_{i}-\sum_{1}^{N}p_{i})]$ in Eq.~(\ref{gN})
simply needs to be replaced by the factor
$\exp[ix(\sum_{0}^{N}k_{i}-\sum_{1}^{N}p_{i})-it(\sum_{0}^{N}k^{2}_{i}-\sum_{1}^{N}p^{2}_{i})]$.
Application of this formalism to non-equilibrium problems is
also straightforward: one has to decompose the initial state, written
in terms of bosonic $\Psi(x)$-operators, into a series of
Zamolodchikov -Faddeev operators $R(\xi)$, generated by the inverse
transform, and then find the time evolution according to
(\ref{Rbasis}) using the direct transform.

Strictly speaking, this expansion is valid only for
an infinite interval. For finite intervals it  gives rise to singularities in
expressions for the correlation functions, which, however, can be corrected by
careful consideration of expressions in each order \cite{CTW}. This
series can be summed up explicitely only for the infinite value of
the coupling $c$ \cite{grosse},\cite{CTW}. Unfortunately this
approach is very difficult to use for calculating time dependence of the
correlation functions at finite $c$ even in equilibrium. We are not
aware of the reproducibility of the asymptotic results which
would correspond to the Luttinger liquid power-laws. On the other
hand the advantage of this approach is the possibility to generalize it
to other specific boundary or initial conditions. In principle this formalism should allow
to include
impurities \cite{crampe} and can be extended to multi-component
generalizations of the NSE \cite{FRS}.

\subsection{Intertwining operator}

Some integrable (and sometime quasi-exactly solvable) models have
the following property
\beq\label{intertwin}
IH_{0}=H_{c}I
\eeq
where two {\it different} Hamiltonians $H_{0,c}$ are connected
(intertwined) by the action of some operator $I$, called the {\it
intertwining} operator. In general, if $H_{0}$ and $H_{c}$ belong to
two different representations of the same algebra, then intertwining
operators exchange these two representations of the same algebras
(or, saying mathematically, establish a certain homomorphism between
them). If  $H_{0}$ is a Hamiltonian for free, {\it noninteracting}
particles, and $H_{c}$ is the Hamiltonian for the {\it interacting}
system with interaction constant given by $c$, then the evolution of
observables in the interacting model can be related to the evolution
of the non-interacting one. Hence {\it dynamics} of the interacting
Hamiltonian can be mapped to dynamics of the noninteracting
Hamiltonian using
\beq
e^{iH_{c}t}=Ie^{iH_{0}t}I^{-1}.
\eeq
Few comments are in order. The existence of the intertwining
operator for integrable models follows from the following facts: (i)
the Bethe states of all integrable models are parametrized by the
integer numbers which label the eigenfunctions of noninteracting
problem. Therefore the wavenumbers of interacting model are
analytically connected to the wavenumbers of the noninteracting one;
(ii) in the language of the coordinate Bethe ansatz a system of $N$
particles is described by the wave function defined in the
$N$-dimensional space divided by the hyperplanes on which the
collision processes occur. Outside these hyperplanes a system
behaves as noninteracting. Transition amplitudes between $N!$
different regions (outer space of the hyperplanes) are the same for
all eigenstates because of the permutation symmetry. These
transition amplitudes define a unitary transformation which is
nothing but the intertwining operator. However, one should note that
in general the operator $I$ is not unitary. Thus, the wavefunctions
of exactly solvable models are not orthonormalized, their overlaps
are given by the determinants of some matrix via Gaudin-Slavnov
formula \cite{gaudin},\cite{BIK}. Defining orthonormalized
wavefunctions one can construct a unitary version of the operator
$I$.

\subsubsection{Degenerate affine Hecke algebra: first quantized notations\label{sec:hecke}}
In the case of the Lieb-Liniger model the intertwining operator can
be related to the representation of the degenerate affine Hecke
algebra \cite{Hikami}. For the XXZ spin chains there should exist
similar intertwining operator between different representations of
the $U_{q}(sl_{2})$ and Temperley-Lieb algebras. One reason to
expect this is because both LL model and XXX spin chain share
essentially the same $R$-matrix satisfying the Yang-Baxter equation.
Some earlier discussion on the XXZ chain are contained in \cite{G3}.

The construction of $I$ for the 1D Bose gas relies on the notion of
Dunkl operator
\beq
\hat{d}_{i}=-i\partial_{i}+i\frac{c}{2}\sum_{j<i}(\epsilon(x_{i}-x_{j})-1)s_{i,j}+i\frac{c}{2}\sum_{j>i}(\epsilon(x_{i}-x_{j})+1)s_{i,j}
\eeq
where $\epsilon(x)$ is a signature function, operator $s_{i,j}$
provides a representation of the Artin's relations for the braids
and exchanges coordinates of particles $i$ and $j$ and satisfy
$x_{i}s_{i,j}=s_{i,j}x_{j}$. The operators ${\hat d}_{i},s_{j}\equiv
s_{j,j+1}$ form a representation of the degenerate affine Hecke
algebra. Another representation of the same algebra is formed by the
ordinary differential (difference) operator $-i\partial_{i}$ and the
integral operator $Q_{i}$ \cite{G2} representing scattering matrix
and acting on the arbitrary function $f(\ldots,x_{i},x_{i+1},\ldots
)$ as
\beq
Q_{i}f(\ldots,x_{i},x_{i+1},\ldots )=f(\ldots,x_{i+1},x_{i},\ldots
)-c\int_{0}^{x_{i}-x_{i+1}}f(\ldots,x_{i}-t,x_{i+1}+t,\ldots )dt
\eeq
The intertwining operator $I$ interchanges these two representations
of the affine Hecke algebra, $(\hat{d}_{i},s_{i,j})$ and
$(-i\partial_{i}, Q_{i})$ and thus intertwines the Dunkl operator
$\hat{d}_{i}$ and the ordinary partial differential operator.
Explicitly
\beq
I=\sum_{w\in S_{N}}\theta(x_{w^{-1}(1)}<\ldots
x_{w^{-1}(N)})s_{w^{-1}}Q_{w}
\eeq
where $s_{w^{-1}}=s_{i_{p}}\ldots s_{i_{2}}s_{i_{1}}$ and
$Q_{w}=Q_{i_{1}}Q_{i_{2}}\ldots Q_{i_{p}}$ ($1\leq
i_{1},i_{2},\ldots,i_{p}\leq N-1$) and where $w$ is a transposition
from the symmetric group $S_{N}$ \cite{Hikami}.

All conserved quantities ${\cal I}_{n}$ for the 1D Bose gas system
are given by the powers of the Dunkl operator, ${\cal I}_{n}
=\sum_{i}\pi(\hat{d}^{n}_{i})$ where $\pi(.)$ is projection onto
symmetric subspace. Thus, the Hamiltonian (\ref{Hc}) is just equal
to ${\cal I}_{2}$.

As a side remark we note that these facts are convenient for the
formulation of the generalized Gibbs ensemble (GGE) approach to
non-equilibrium dynamics of the LL model \cite{GGE}. In particular,
the GGE density matrix of the 1D Bose gas should be related to the
noninteracting density matrix by the intertwining relation. As
discussed in Ref.~\cite{KE}, the GGE conjecture should work when
eigenvalues of the integrals of motion are parametrized by either
$\{0,1\}$ (fermionic-like systems) or by integers $\{0,1,2,\ldots\}$
(bosonic-like systems). The case of the Bose gas belongs to the
second class. We therefore conjecture that the GGE is applicable to
the 1D Bose gas (LL model), at least for the local observables.

\subsubsection{Graphical representation: Gutkin approach\label{sec:gutkin}}
An interesting approach to the 1D Bose gas based on the intertwining
operator has been developed by E. Gutkin in a series of papers
\cite{G3},\cite{G2},\cite{G1},\cite{DG}. It was later applied to the
problem of propagation of an optical pulse in nonlinear media in
Ref.~\cite{PY}.

In this approach the intertwining operator is expanded as a series
of different contributions labeled by the so-called collision
graphs~\footnote{The graph $\Gamma$ with $q$ vertices and $p$
oriented edges is called collision graph if a) for every vertex
there is an edge coming into it or going out of it; b) there is at
most one edge between any two vertices; c) all vertices can be
labeled in an ordered way from 1 to $q$ such that edges go from
smaller to larger number. }. The evolution of the quantum field
$\Psi(x,t)$ governed by the NSE is given entirely in terms of the
evolution of the field $\Psi_{0}(x,t)$ for the free Schr\"{o}dinger
equation,
$i\partial_{t}\Psi_{0}(x,t)=-\partial_{x}^{2}\Psi_{0}(x,t)$, and by
the quantities $a_{\Gamma}$ (defined in terms of some distributions)
entering the expansion of the intertwining operator over a set of
collision graphs
\beq
I=\sum_{\Gamma,n=q(\Gamma)}\int d^{n}x\int d^{n}y
a_{\Gamma}(x_{1},x_{2},\cdots,x_{n},y_{1},y_{2},\cdots
y_{n})\Psi_{0}^{\dag}(x_{1})\Psi_{0}^{\dag}(x_{2})\cdots\Psi_{0}^{\dag}(x_{n})\Psi_{0}(y_{1})\Psi_{0}(y_{2})\cdots\Psi_{0}(y_{n})
\eeq
and similarly for the operator $I^{-1}$ with $a_{\Gamma}$ replaced
by the related distribution $b_{\Gamma}$. There is a special choice
of $a_{\Gamma}$ and $b_{\Gamma}$ for which the operator $I$ is
unitary. For details about this approach we refer to the original
papers by Gutkin.

The advantage of this approach is the possibility to find the
propagator of the nonlinear problem and thus, potentially, solve the
initial state evolution problem for an arbitrary initial state.
Computationally one only needs to deal with the free fields
$\Psi_{0}(x)$ and their time evolution with a free (noninteracting)
Hamiltonian. We note that the collision graph expansion is highly
non-perturbative and can not be rewritten as an expansion in powers
of $c$ in opposition to the Gelfand-Levitan-based expansion of
Sec.~\ref{sec:inv-scatt}. A disadvantage of this approach is rapidly
growing complexity of collision graph expansion with increasing
number of particles. Also it is not known at present whether some of
the most relevant graphs can be summed up in order to develop useful
approximation schemes.

\subsubsection{Second quantized approach\label{sec:second_quant}}
In series of papers \cite{SK} Sasaki and Kebukawa developed a
field-theoretical approach to the LL model. Although they did not
present it in this context, the second quantized form of the
intertwining operator can be recognized in their construction
\footnote{It is also interesting to note that later on the same
authors constructed similar second-quantized approach to the
fermionic Yang-Gaudin model as well}. In this approach one considers
the LL Hamiltonian written in the usual second-quantized form,
\beq
H_{c}=\sum_{p}\frac{p^{2}}{2m}a_{p}^{\dag}a_{p}+\sum_{p,q,r}\frac{c}{2L}a^{\dag}_{p+r}a^{\dag}_{q-r}a_{q}a_{p}.
\eeq
Here $a_{p}^{\dag},a_{p}$ are bosonic creation and annihilation
operators corresponding to momenta $p_{i}=2\pi\hbar n_{i}/L$
($n_{i}$ should be integer). The second-quantized analog of the
Bethe ansatz wavefunction for $N$ particles has the form
\beq
|\Psi_{q_{1},q_{2},\ldots q_{N}}\rangle=B_{q_{1},q_{2},\ldots
q_{N}}\sum_{p_{i,j};1\leq i<j\leq N}\prod_{1\leq i<j\leq
N}d(p_{i,j};k_{i,j})\prod_{i=1}^{N}a^{\dag}_{\sum_{j=1,j\neq
i}^{N}p_{i,j}+q_{i}}|0\rangle
\eeq
where $B_{q_{1},q_{2},\ldots q_{N}}$ is a normalization factor,
$p_{i,j}=p_{i}-p_{j}=-p_{j,i}$ and $q_{i}$ ($i=1,\ldots, N$) are
defined as an integer$\times 2\pi\hbar/L$,
$d_(p_{i,j};k_{i,j})=-k_{i,j}/(p_{i,j}-k_{i,j})$ and
$k_{i,j}=-k_{j,i}$ are solutions of the Bethe ansatz equations
\beq
\cot(\frac{Lk_{i,j}}{2\hbar})=\frac{\hbar}{mc}(k_{i}-k_{j})=\frac{\hbar}{mc}(2k_{i,j}+\sum_{l\neq
i,j}(k_{i,l}-k_{j,l})+q_{i}-q_{j}),\qquad 1\leq i<j\leq N.
\eeq
The eigenvalues in these notations are given by
$E_{q_{1},\ldots,q_{N}}=\sum_{i=1}^{N}k_{i}^{2}/2m=\sum_{i=1}^{N}\sum_{j\neq
i}(k_{i,j}+q_{i})^{2}/2m$.

It is interesting that authors of \cite{SK} found a {\it unitary}
operator which transforms eigenstates of noninteracting system
$H_{0}$ into eigenstates of the interacting Hamiltonian $H_{c}$. In
the second quantized form this unitary operator is given by
\beq
U=\sum_{q_{1}\leq q_{2}\leq\ldots q_{n}}\sum_{p_{i,j};1\leq i<j\leq
n}B_{q_{1},q_{2},\ldots,q_{n}}A_{q_{1},q_{2},\ldots,q_{n}}\prod_{1\leq
i<j\leq n}d(p_{i,j};k_{i,j})\prod_{i=1}^{n}a^{\dag}_{\sum_{j=1;j\neq
i}^{n}p_{i,j}+q_{i}}\prod_{i=1}^{n}a_{q_{i}}
\eeq
where $A_{q_{1},\ldots,q_{N}}$ are normalization factors for
non-interacting eigenstates. Explicit calculations show that this
unitary operator is indeed equivalent to the intertwining operator
from previous subsections.

This approach has advantages for initial states which can be readily
represented using formalism of second quantization. For the case of
a large coupling constant it gives the same result as the one used
later on in the text.

\subsection{q-bosons}
In a series of recent papers \cite{BIKq},\cite{Kundu} a system of
$q$-bosons hopping on 1D lattice has been studied. This model is
defined by its Hamiltonian
\begin{equation}
H_q=-\frac{1}{2} \sum_{n=1}^{M} (b_n^{\dagger}b_{n+1}+b_n
b_{n+1}^\dagger - 2 N_n) \label{qHam}
\end{equation}
where the periodic boundary condition is imposed. Operators $B_n$,
$B_n^\dagger$ and $N_n$ satisfy the $q$-boson algebra
\beq
~[N_i, b_j^\dagger]=b_j^\dagger \delta_{ij},  \qquad [N_i,
b_j]=-b_i\delta_{ij}, \qquad [b_i, b_j^\dagger]&=q^{-2
N_i}\delta_{ij} \label{qAlg}
\eeq
where $q=e^{\gamma}$. The operator of the total number of particles
$N=\sum_{j=1}^{M} N_j$ commutes with the Hamiltonian $H_q$.

The continuum limit of the model is defined using the limiting
procedure
\begin{equation}
\delta\to 0, \qquad M\delta=L, \qquad \gamma=\frac{c \delta}{2}.
\label{qCont}
\end{equation}
When applied to the system of $q$-bosons this procedure gives
\begin{equation}
H=\int_0^{L} dx \left[\partial_x b^{\dagger}(x) \partial_x b(x)+ c
b^\dagger(x)b^\dagger(x) b(x)b(x)\right],
\label{qContHam}
\end{equation}
where $[b(x), b^{\dagger}(y)]=\delta(x-y)$ are the canonical Bose
fields. Therefore 1D Bose gas with contact interaction can be
regarded as a continuum limit of the $q$-boson lattice model.
Apparently a solution of the inverse scattering problem for the NSE
should be related to the one for the $q$-boson model. An explicit
form of this relationship is not known, although should
exist~\footnote{The authors are grateful to M. Zvonarev for
discussions on this issue.}.

This approach can be used to study non-equilibrium dynamics.
Solutions of particular evolution problem for $q$-bosons can be
transformed into solutions of NS problem via the limiting procedure
(\ref{qCont}). An attempt to construct the evolution operator for a
single-mode $q$-boson system has been made in \cite{CD}. Even for
this simple case the $q$-path integral has an involved structure
which includes integration with the measure corresponding to
non-flat phase space and complicated action having non-trivial
dynamical phase. All this makes the possibility of explicit
evaluation of the $q$-path integral questionable. It is interesting
to note that dynamics in this phase space is naturally constrained
by integrals of motion. This is reminiscent of recent suggestions of
the importance of GGE for dynamics of integrable models \cite{GGE}.

\subsection{Extended conformal symmetry\label{sec:ext-symm}}

It is known that the low-energy description of the 1D Bose gas can
be formulated as a bosonic Luttinger liquid
\cite{Haldane,Cazalilla}. This description operates with a linear
spectrum of bosonic modes and treats the interaction part
essentially exactly. From the more formal point of view, in this
low-energy description the thermodynamic limit of the NS system
represents a conformal field theory (CFT) with the central charge
$c=1$. This means, in particular, that the low-energy (bosonic
Luttinger) Hamiltonian is a linear combination of the zero-momentum
generators $L_{0},\bar{L}_{0}$ (for the right and for the left
moving parts) of the conformal symmetry algebra, i.e. the Virasoro
algebra
\beq
[L_n , L_m ]=(n-m) L_{n+m}+\frac{1}{12}n(n^{2}-1)\delta_{n+m,0}
\eeq
and the same for $\bar{L}_{n}$, ($[L_{n},\bar{L}_{m}]=0,\forall
n,m$). The space of states is the so-called Verma module
(representation space) for this Virasoro algebra. The Kac table
defines a conformal tower for this $c=1$ CFT which allows to compute
any correlation functions. One could also include finite size
corrections and effects of irrelevant perturbations
\cite{BM},\cite{Berkovich}.

The non-equilibrium dynamics can not be correctly described in terms
of the low-energy, Luttinger liquid theory only. The reason is
two-fold: first, if the initial state represents a "large"
perturbation over equilibrium, which includes the overlap with the
whole spectrum, the excitations involving nonlinear part of the
spectrum should be important. Second, even for relatively weak
perturbations, which do not thermalize because of integrability, the
dynamics at large times leads to growing correlations over the
entire momentum space. So irrelevant contributions from the
interaction part of the Hamiltonian should become important at long
times.

It appears that nonlinearity of the dispersion as well as irrelevant
(from the equilibrium point of view) parts of interaction can be
included using the {\it extended} conformal symmetry called ${\cal
W}_{1+\infty}$. The ${\cal W}_{1+\infty}$ algebra is a
representative of a class of nonlinear algebras which appeared in
conformal field theories. It is a product of a quantum version of an
algebra of area-preserving diffeomorphisms of a 2D cylinder and the
abelian Kac-Moody algebra. The presence of dynamical symmetry
 ${\cal W}_{1+\infty}$ is a general property of
gapless one-dimensional systems. The chiral generators
$W_{n}^{\alpha}$ of ${\cal W}_{1+\infty}$ are labeled by the
momentum index $n, k=2\pi n/L$ and by the conformal spin index
$\alpha =0,1,2,\ldots\infty$. They satisfy
\beq
~[W_{n}^{\alpha},W_{m}^{\beta}]=(\beta n -\alpha
m)W^{\alpha+\beta-1}_{n+m}+q(\alpha,\beta,n,m)W^{\alpha+\beta-3}_{n+m}+\ldots+\delta^{\alpha\beta}\delta_{n+m,0}c
d(\alpha, n)
\eeq
where $q(\alpha,\beta,n,m)$ and $d(\alpha,n)$ are known polynomials
of their arguments \cite{winf}.

For many interesting cases the Hamiltonian can be written as a
linear and bilinear combinations of Cartan generators of the ${\cal
W}_{1+\infty}$ algebra\footnote{Cartan generators of any algebra are
those which commute with each other. In our case it means that
$[W^{\alpha}_{0},W^{\beta}_{0}]=0,\forall\alpha,\beta$}. The
representation theory of $W_{1+\infty}$ in the Verma module can be
constructed in the same way as representation theory for the
Virasoro algebra \cite{KacRadul} starting from the highest weight
state and building a towers of descendant states. Using this
representation theory one can construct the action of ${\cal
W}_{1+\infty}$ generators on bosonic Fock space. One can therefore
establish a correspondence between Bethe ansatz eigenstates and
certain combination of states of representation module of the ${\cal
W}_{1+\infty}$. The evolution operator is factorized in the product
of factors corresponding to different Cartan generators
\beq
U(t) =e^{iHt}=\prod_{\alpha}e^{i\beta_{\alpha}W^{\alpha}t}
\eeq
where $\beta_{\alpha}$ are functions of the interaction strength and
a level. When the initial state can be  expressed as a linear
combination of descendant states, the application of $U(t)$ is
straightforward.

In practice it is difficult to deal with the whole nonlinear
dispersion, so, for practical purposes one can keep only
next-to-linear terms in this expansion. In this case the Hamiltonian
will include only a finite number of ${\cal W}_{\infty}$ generators.
For example, in the strong coupling regime, corrections to the
Hamiltonian to the first order in the inverse power of the
interaction strength,
\beq
H=\frac{2\pi\rho}{LK}(W^{1}_{0}+\bar{W}^{1}_{0})+\frac{2\pi}{L^{2}K}[W^{2}_{0}+\bar{W}^{2}_{0}]+\ldots
\eeq

The advantage of this approach is simplicity of including
higher-order corrections originating from  nonlinearity of
dispersion, and the potential to solve the problem exactly. The
disadvantage of this method comes from its limitation to treat
gapless systems only. Details of this approach together with several
related questions will be presented in a separate work
\cite{VGwinf}.

\subsection{Form factors and decomposition of the initial state\label{sec:decomp}}

This is the most direct approach which we will use further in this
paper. In this approach we decompose the initial state in terms of
the eigenstates of the Hamiltonian. The latter form a complete and
orthogonal set of states. To compute time evolution of the
correlation functions we have to combine several ingredients: (i)
find complete basis of many body wave functions; (ii) know exact
eigenvalues; (iii) determine matrix elements (or {\it form-factors})
of various operators in the basis of exact eigenfunctions; (iv) find
decomposition of the initial state in the complete set of exact
many-body states; (v) develop effective procedure for summation over
intermediate states.

Ingredients (i)-(ii) are known from the Bethe ansatz exact solution.
Matrix elements (iii) were computed in many exactly soluble models
on the basis of the so-called determinant representation. Recent
progress in computation of the matrix elements of operators in the
basis of the Bethe Ansatz wave functions makes it possible to
advance in this direction. Decomposition of the initial states (iv)
over the complete basis depends significantly on the concrete nature
of the states. We were able to evaluate these overlaps for a 1D Bose
gas at large interaction strength for two types of initial
conditions: all particles in a zero momentum state and all particles
in Gaussian wavefunction in real space. Currently we perform
summation over intermediate states numerically. This imposes certain
restrictions on the number of particles in our system. We checked
that in equilibrium this number is sufficient to saturate any
correlators to their values in the thermodynamic limit. To
illustrate this observation we plot the $g_{2}(x)$, computed in the
ground state using our method, and compare it to the known analytic
expression in Fig.(\ref{fig.ground.spatial}).

Here we make general comments on the structure of the phase space of
form-factors. For a systems with a gap only a small number of states
needs to be taken into account. Contributions of many-particle
states are suppressed by the gap. This is the case for the NLS
system with attraction, which can be identified with the
nonrelativistic limit of the quantum sine-Gordon model for which
previous statement is also correct (see e.g. review \cite{KE} and
refs. therein for direct comparison of different contribution to the
form-factor expansion of the correlation functions). Contributions
coming from higher order terms grow in the limit of long time
evolution, but should only generate subleading corrections. The
situation is very different for gapless systems, such as the NLS
system with repulsion. Many-particle contributions are suppressed at
most as a power law. Hence ideally we need to take into account the
entire phase space. This makes the problem of summation over
intermediate states very difficult. In this paper we deal with the
most difficult case of repulsively interacting Bose gas in the
regime of strong interaction, when, in principle, the allowed phase
space is huge and multi-particle states are in general not
suppressed.

The form-factor approach has been successfully applied to
computation of equilibrium correlation functions in many models
\cite{smirnov}. A review of the applications of this method to
massive models is given in Ref.~\cite{EK}. The gap in the spectrum
leads to the rapid convergence of the form-factor expansion. It is
enough to take contributions from a few particles to saturate any
correlation function. This is not the case for massless models where
contributions of multi-particle processes are essential. Recently
this approach was applied to compute equilibrium correlation
functions of 1D Bose gas in \cite{caux06},\cite{CCS}, (see also
Ref.~\cite{CCattr} for the attractive case~\footnote{Note that
direct quantum-mechanical computations has been done in
Ref.~\cite{LH}; they agree with \cite{CCattr} up to multiplicative
factor.}). Earlier studies reviewed in \cite{BIK} allowed to
evaluate various asymptotics of the time-dependent correlation
functions at equilibrium (for concrete example of non-zero
temperature see e.g. \cite{KS3},\cite{KS2}.) Recently, a
form-factor approach has been applied to time evolution of specific
initial state in the XXZ model \cite{MC}. Specific choice of the
initial state made it possible to perform analytical
analysis of the time-evolved. This is one of the rare examples
when a work function and Loschmidt echo, important quantities
related to the properties of a stationary state \cite{Silva}, have
been evaluated analytically.

One of the advantage of the direct application of the form-factor
technique is the possibility to consider non-equilibrium dynamics of
weakly non-integrable systems. In Refs.~\cite{DMS} the form-factor
perturbation theory has been developed for a class of models which
deviate weakly from an integrable "fixed point". This perturbation
theory is unusual because its unperturbed states are highly
correlated states of interacting integrable theory. We expect that
this formalism can be extended to treat non-equilibrium dynamics as
well.


The form-factor approach supplemented by a procedure of decomposing
the initial state into a complete set of many-body eigenstates is
universal and can be applied to a large class of integrable models.
This method allows several extensions and modifications. In the rest
of this paper we focus on this method and demonstrate that it is a
powerful tool for analyzing non-equilibrium evolution of correlation
functions.

In the next section we also show that one can use the intertwining
operator to find an overlap coefficient between interacting and
noninteracting states. We do this explicitly in the strong coupling
limit.\footnote{We also comment on some particular type of models
where these and other overlaps can be found explicitly for various
coupling constants. This is the case for some type of Gaudin models,
e.g. Richardson model of mesoscopic BCS pairing and the central spin
problem. In these models the BA states do not depend explicitly on
the strength of the coupling constant, so one could use the
Gaudin-Slavnov determinant formula for these overlaps.}
To conclude this section we note that other  approaches to
calculating correlation functions in integrable systems
undergoing nonequilibrium dynamics  have been discussed in Refs.~\cite{HY}, and
\cite{IKR}. When initial states can
be written using ordering of the coordinates as in eq.
(\ref{ord_state}) (see Appendix A),  an approach to calculating
their dynamics has been proposed in Ref. \cite{TW}.
We expect that the approach developed in \cite{TW}  is related to the
method which we discuss in the current paper.
Yet another approach, closely related to the
form-factor method in the limit of an {\it infinite volume}, allows for
computation of the time evolution of the one-point function. This
method relies on representing {\it certain} (integrable)
initial states as {\it boundary states}, which correspond to coherent
superpositions of the eigenstates of the model \cite{GDLP,FM}.
Recently, an interesting approach to equilibrium correlators in the
Lieb-Liniger model has been developed in Ref. \cite{KMT-KMP}. The idea of
the latter papers is to consider first a relativistic {\it
massive} model (the so-called sinh-Gordon model), which, in specific
non-relativistic limit, is equivalent to the Lieb-Liniger model. The
form-factors of the sinh-Gordon model are known and the form-factor
expansion of correlation
functions is rapidly convergent. Such convergence
is a result of the mass gap in the spectrum. In
the special non-relativistic limit, all expressions for the
field operators and correlators of the sinh-Gordon model go over
into corresponding expressions for the LL model. Moreover, the
non-relativistic limit benefits from the rapid convergence of the
form-factor expansions in its relativistic counterpart. One might
hope that this interesting approach can be extended to
non-equilibrium dynamics.

\section{Exact approach to non-equilibrium dynamics based on form-factors}

In this section we describe the application of the method of
form-factors to the computation of non-equilibrium correlation
functions. This method is general for any integrable system. In this
paper we focus on the 1D Bose gas. As discussed earlier the
treatment of gapless system is more challenging because of the large
number of contributions from intermediate states which need to be
included. To overcome this difficulty we devised a special numerical
procedure of summation over intermediate states. This procedure is
described in the next section.

\subsection{Formulation of the problem}
To fix notations we consider the nonlinear Schr\"{o}dinger
Hamiltonian on a finite interval $[0,L]$
\beq\label{ham}
H=\int_{0}^{L} dx
[\partial_{x}\Psi^{\dag}(x)\partial_{x}\Psi(x)+c\Psi^{\dag}(x)\Psi^{\dag}(x)\Psi(x)\Psi(x)]
\eeq
where the commutation relation between bosonic field $\Psi(x)$ and
its conjugated $\Psi^{\dag}(x)$ as well as the pseudovacuum state
$|0\rangle$ are defined as \cite{BIK}
\beq
~[\Psi(x),\Psi^{\dag}(y)]&=&\delta(x-y),\qquad
[\Psi(x),\Psi(y)]=[\Psi^{\dag}(x),\Psi^{\dag}(y)]=0\\
\Psi(x)|0\rangle &=&0,\qquad \langle 0|\Psi^{\dag}=0,\qquad \langle
0|0\rangle =1.
\eeq
The total number of particles and the momentum operators are
conserved  quantities
\beq
N&=&\int_{0}^{L}\Psi^{\dag}(x)\Psi(x), \qquad
P=-\frac{i}{2}\int_{0}^{L}[\Psi^{\dag}(x)\partial_{x}\Psi -
(\partial\Psi^{\dag}(x))\Psi(x)]\\
~[H,N]&=&[H,P]=0
\eeq

In each $N$-particle sector this system is equivalent to the 1D Bose
gas with contact interaction potential (Lieb-Liniger model)
\cite{LL}, \cite{McG},\cite{Ber},\cite{gaudin}. In connection to the
previous section we note that inverse scattering transform for this
problem has been developed in many papers (see
\cite{Sklyanin1},\cite{Faddeev},\cite{CTW},\cite{G1} for extensive
reviews).

The problem we are asking here can be formulated as follows: suppose
we prepared a system in a certain initial $N$-particle state
$|\psi_{0}^{(N)}\rangle$ (e.g. coherent state can be considered as a
superposition of states with different $N$). This state evolves
according to the Hamiltonian (\ref{ham}). The question is to compute
the correlation functions of the following type
\beq
&\langle
\psi_{0}^{(N)}|\Psi^{\dag}(x_{1},t_{1})\Psi(x_{2},t_{2})|\psi_{0}^{(N)}\rangle,&\label{psipsi}\\
&\langle\psi_{0}^{(N)}|\Psi^{\dag}(x_{1},t_{1})\Psi(x_{1},t_{1})\Psi^{\dag}(x_{2},t_{2})\Psi(x_{2},t_{2})|\psi_{0}^{(N)}\rangle,&\\
&\langle\psi_{0}^{(N)}|\exp(\alpha Q(x))|\psi_{0}^{(N)}\rangle,&
\eeq
where the last formula expresses a generation function for the
density-density correlator,
\beq
Q(x,t)=\int_{0}^{x}\Psi^{\dag}(y,t)\Psi(y,t)dy
\eeq
The density is defined as
\beq
\rho(x,t)=\Psi^{\dag}(x,t)\Psi(x,t)
\eeq

Assuming that the initial state is normalized, one can insert a
resolution of unity several times,
\beq
{\bf 1}=\sum_{\{\lambda\}}
\frac{|\{\lambda\}\rangle\langle\{\lambda\}|}{\langle\{\lambda\}|\{\lambda\}\rangle}
\eeq
into expression for the correlation functions. This suggests to
define normalized correlation functions,
\beq
& &\langle\psi_{0}^{(N)}|O(x_{1},t_{1})O(x_{2},t_{2})|\psi_{0}^{(N)}\rangle \nonumber\\
&=&\sum_{\{\lambda\}_{N}}\sum_{\{\mu\}_{N}}\sum_{\{\nu\}_{N}}\frac{\langle
\psi_{0}^{(N)}|\{\lambda\}_{N}\rangle\langle\{\lambda\}_{N}|
O(x_{1},t_{1})|\{\mu\}_{N}\rangle\langle\{\mu\}_{N}|O(x_{2},t_{2})|\{\nu\}_{N}\rangle\langle\{\nu\}_{N}|\psi_{0}^{(N)}\rangle}
{\langle\{\lambda\}|\{\lambda\}\rangle\langle\{\mu\}|\{\mu\}\rangle\langle\{\nu\}|\{\nu\}\rangle}.\label{rhorho}
\eeq
In the following we concentrate on the density-density correlation
function. According to our programme we need matrix elements of the
operators in the complete basis of states. In this basis time and
space dependence are trivial: using the Galilei invariance (we
implicitly assume periodic boundary conditions) the matrix elements
which we need are given by the expressions
\beq
\langle\{\mu\}_{N}|Q(x,t)|\{\lambda\}_{N}\rangle=(e^{ix(P_{\lambda}^{(N)}-P_{\mu}^{(N)})}-1)e^{it(E_{\lambda}^{(N)}-E_{\mu}^{(N)})}
F_{Q}(\{\mu\};\{\lambda\}).
\eeq

Here the energy and momentum are given by
\beq\label{enmom}
E_{\lambda}^{(N)}=\sum_{j=1}^{N}\lambda_{j}^{2},\qquad
P_{\lambda}^{(N)}=\sum_{j=1}^{N}\lambda_{j}.
\eeq
In the following it will be convenient to use more compact notations
\beq
A_{\lambda}^{(0)}&=&\langle\psi_{0}^{(N)}|\{\lambda\}_{N}\rangle\\
F_{Q}(\{\mu\};\{\nu\})&=&\langle\{\mu\}_{N}|Q(0,0)|\{\nu\}_{N}\rangle\\
||\{\lambda\}||&=&\langle\{\lambda\}|\{\lambda\}\rangle.
\eeq

We can write down an expression for the density-density correlator
as
\beq
&&\langle\psi_{0}^{(N)}|\rho(x,t)\rho(0,0)|\psi_{0}^{(N)}\rangle\\
&=&\frac{\partial^{2}}{\partial x_{1}\partial
x_{2}}\sum_{\{\lambda\}_{N}}\sum_{\{\mu\}_{N}}\sum_{\{\nu\}_{N}}
\frac{A_{\lambda}^{(0)}(A_{\nu}^{(0)})^{*}F_{Q}(\{\lambda\};\{\mu\})(F_{Q}(\{\mu\};\{\nu\}))^{*}}{||\{\lambda\}||||\{\mu\}||||\{\nu\}||}\nonumber\\
&\times&(e^{ix_{1}(P_{\mu}^{(N)}-P_{\lambda}^{(N)})}-1)e^{it_{1}(E_{\mu}^{(N)}-E_{\lambda}^{(N)})}
(e^{ix_{2}(P_{\nu}^{(N)}-P_{\mu}^{(N)})}-1)e^{it_{2}(E_{\nu}^{(N)}-E_{\mu}^{(N)})}
\label{eq.rho_rho}
\eeq
where $*$ means complex conjugation.

\subsection{Bethe Ansatz Ingredients for 1D Bose gas}
As discussed above in Section (\ref{sec:decomp}) we need several
ingredients: a set of exact eigenstates and eigenenergies of the
model; an overlap of the initial state with the eigenstates; an
explicit expressions for the matrix elements of operators in the
eigenbasis.

\subsubsection{BA states} The BA states are described by a set of $N$
real numbers $\lambda$ which are given by the solution of the BA
equations \cite{LL},\cite{gaudin},\cite{BIK}
\beq\label{BAeq}
\lambda_{j}+\frac{1}{L}\sum_{l=1}^{N}2\arctan\frac{\lambda_{j}-\lambda_{l}}{c}=\frac{2\pi}{L}(I_{j}-\frac{N+1}{2}),\quad
j=1\ldots N
\eeq
Here the quantum numbers $I_{j}$ are half-odd integers for $N$ even,
and integers for $N$ odd. For eigenfunctions rapidities do not
coincide, $\lambda_{j}\neq\lambda_{k}$ for $j\neq k$. The whole Fock
space is obtained by choosing sets of ordered quantum numbers
$I_{j}>I_{k}$, $j>k$ meaning that $\lambda_{j}>\lambda_{k}$, $j>k$.

From solution of these equations we immediately obtain energies and
momenta via Eq.(\ref{enmom}).

\subsubsection{Overlaps between BA states} The overlap between BA
states is given by the Gaudin-Korepin formula \cite{K1}, \cite{S1}
\cite{gaudin}, \cite{BIK},\cite{K3},
\cite{K2},\cite{S2},\cite{KS},\cite{CCS}
\beq\label{Gaudin}
\langle\{\lambda\}_{N}|\{\lambda\}_{N}\rangle=c^{N}\prod_{j>k=1}^{N}\frac{(\lambda_{j}-\lambda_{k})^{2}+c^{2}}{(\lambda_{j}-\lambda_{k})^{2}}\det_{N}{\cal
G}(\{\lambda\})
\eeq
where
\beq
{\cal
G}_{jk}(\{\lambda\})&=&\delta_{jk}\left[L+\sum_{l=1}^{N}K(\lambda_{j},\lambda_{l})\right]-K(\lambda_{j},\lambda_{k})\\
K(\lambda_{j},\lambda_{k})&=&\frac{2c}{(\lambda_{j}-\lambda_{k})^{2}+c^{2}}\label{k}
\eeq



\subsubsection{Matrix elements} The matrix elements (form-factors) for the current
operator are given by the following determinantal expression
\cite{S3},\cite{KS}, \cite{CCS}
\beq
F_{Q}(\{\mu\}_{N};\{\lambda\}_{N})&=&
\frac{i^{N}}{(V_{N}^{+}-V_{N}^{-})}\prod_{j,k=1}^{N}
(\frac{\lambda_{j}-\lambda_{k}+ic}{\mu_{j}-\lambda_{k}})\nonumber\\
&\times&\det\left(\delta_{jk}(\frac{V_{j}^{(+)}-V_{j}^{(-)}}{i})+\frac{\prod_{a=1}^{N}(\mu_{a}-\lambda_{j})}{\prod_{a\neq
j}^{N}(\lambda_{a}-\lambda_{j})}[K(\lambda_{j}-\lambda_{k})-K(\lambda_{N}-\lambda_{k})]\right)
\label{eq.F}
\eeq
where
\beq
V_{j}^{(\pm)}=\frac{\prod_{a=1}^{N}(\mu_{a}-\lambda_{j}\pm
ic)}{\prod_{a=1}^{N}(\lambda_{a}-\lambda_{j}\pm ic)}
\label{eq.Vjpm}
\eeq
and where $K$ is given by (\ref{k}). The matrix inside $\det()$ has
only real entries. This matrix has size $N\times N$. Real numbers
$\{\lambda\},\{\mu\}$ are solutions of BA equations (\ref{BAeq}).

\subsection{Overlaps of the initial states with Bethe Ansatz basis}
Here we study in details two physically motivated examples of
initial states. The first one has its origin in the condensate
physics whereas the second one describes a Gaussian pulse created in
special nonlinear media. Before going to these examples we make
several general observations.

\subsubsection{General remarks} To compute the overlap with the
initial state we consider the following (coordinate) representation
for the Bethe ansatz states of the NSE
\beq
|\Psi_{N}(\lambda_{1},\ldots\lambda_{N})\rangle
=\frac{1}{L^{N}\sqrt{N!}}\int d^{N}z
\chi_{N}(z_{1},\ldots,z_{N}|\lambda_{1},\ldots,\lambda_{N})\Psi^{\dag}(z_{1})\ldots\Psi^{\dag}(z_{N})|0\rangle
\eeq
where the function $\chi_{N}$ is given by
\beq
\chi_{N}(z_{1},\ldots,z_{N}|\lambda_{1},\ldots,\lambda_{N})=const\prod_{N\geq
j>k\geq 1}(\frac{\partial}{\partial z_{j}}-\frac{\partial}{\partial
z_{k}}+c)\det[\exp(i\lambda_{j}z_{k})]\nonumber\\
=\frac{1}{\sqrt{N!\prod_{j>k}[(\lambda_{j}-\lambda_{k})^{2}+c^{2}]}}\sum_{{\cal
P}}(-1)^{[{\cal P}]}\exp[i\sum_{n=1}^{N}z_{n}\lambda_{{\cal
P}n}]\prod_{j>k}[\lambda_{{\cal P}j}-\lambda_{{\cal
P}k}-ic\epsilon(z_{j}-z_{k})].\nonumber\\
\label{eq.chi.dres}
\eeq
It is convenient to define the basic wave function
\beq
|\Psi\rangle_{0}^{(N)}=\frac{1}{\sqrt{N!}}\Psi^{\dag}(x_{1})\ldots\Psi^{\dag}(x_{N})|0\rangle.
\eeq
In principle, any quantum state can be expanded as
\beq
|\Psi^{(0)}\rangle
=\sum_{n}a_{n}\int\frac{1}{\sqrt{n!}}f_{n}(x_{1},\ldots,x_{n})\Psi\rangle_{0}^{(n)}
\eeq
with
\beq
\sum_{n}|a_{n}|^{2}=1,\qquad \int
|f_{n}(x_{1},\ldots,x_{n})|^{2}dx_{1}\ldots dx_{n}=1
\eeq

Intuitively one expects that the evolution of some {\it special}
initial states can, under certain assumptions about these states and
about the Hamiltonian, be relatively simple. This is indeed the case
for the 1D Bose gas. Using the expansion (\ref{glmexp}) and contour
integration it was shown in \cite{CTW} that the state defined as
\beq\label{eq:initial}
|\psi_{N}(t=0)\rangle_{ord}=\theta(x_{1}>x_{2}>\ldots>x_{N})|\Psi^{\dag}(x_{1})\Psi^{\dag}(x_{2})\ldots\Psi^{\dag}(x_{N})|0\rangle
\eeq
is equal to the state $R^{\dag}(x_{1})R^{\dag}(x_{2})\ldots
R^{\dag}(x_{N})|0\rangle$ where
$R(x)=\int\frac{d\xi}{2\pi}e^{ix\xi}R(\xi)$ (see Section
\ref{sec:inv-scatt} for definitions). The calculation of the
evolution of this state is therefore straightforward.

\subsubsection{The overlap with a delta function in momentum space.}

We first construct a Fourier transform of the basic state:
\beq
|\Psi(q_{1}),\ldots\Psi(q_{N})\rangle =
\frac{1}{\sqrt{N!L^{N}}}[\int_{0}^{L}e^{iq_{1}x_{1}}\Psi^{\dag}(x_{1})
dx_{1}],\ldots[\int_{0}^{L}e^{iq_{N}x_{N}}\Psi^{\dag}(x_{N})
dx_{N}]|0\rangle.
\eeq
At the end of all computations one could take all momenta to be
equal $q_{1}=q_{2}=\ldots q_{N}$ to obtain a condensate state
\beq
|\psi^{(N)}_{0}(q)\rangle_{C} = [\Psi^{\dag}(q)]^{N}|0\rangle.
\eeq

Let us consider the overlap of this state with the {\bf
Tonks-Girardeau} wave function corresponding to the large $c$ limit.
There are several reasons to focus on this limit: 1) from our basic
interest in application to quantum optics with strongly-correlated
photons the TG case is the most interesting since effects of
fermionizations \cite{MG} should be most pronounced \cite{Chang} in
this limit; 2) equations for the overlap are relatively simple and
amenable to analytical calculations; 3) one could use an expansion
in powers of $1/c$ around TG limit to produce results for the
correlation functions beyond TG limit.

In the TG limit the wave function takes the following form
\cite{TG}, \cite{S4},\cite{kojima}:
\beq
|\Psi_{TG}\rangle=\frac{1}{\sqrt{N!}}\int_{0}^{L}dx_{1}\ldots
\int_{0}^{L}dx_{N}\chi^{(N)}_{TG}(\{\lambda_{i}\}|\{x_{i}\})\Psi^{\dag}(x_{1})\ldots\Psi^{\dag}(x_{N})|0\rangle
\eeq
such that $\langle\Psi_{TG}|\Psi_{TG}\rangle=1$ and where
\beq
\chi^{(N)}_{TG}(\{\lambda_{i}\}|\{x_{i}\}) =
\frac{1}{\sqrt{N!L^{N}}}\prod_{1\leq j<k\leq
N}\epsilon(x_{j}-x_{k})\det_{N}(e^{i x_{m}\lambda_{n}})
\eeq
Here
\beq
\lambda_{i}=\frac{2\pi}{L}n_{i},\qquad \left\{
                                   \begin{array}{ll}
                                     n_{i}\in {\bf Z}, & N\quad\mbox{odd;} \\
                                     n_{i}\in {\bf Z}+\frac{1}{2}, & N\quad\mbox{even.}
                                   \end{array}
                                 \right.
\eeq
The overlap therefore is reduced to the following integral:
\beq
_{C}\langle
\psi^{(N)}_{0}(q)|\Psi_{TG}\rangle=\frac{N!}{N!L^{N}}\int_{0}^{L}dx_{1}\int_{0}^{L}dx_{2}\ldots\int_{0}^{L}dx_{N}\prod_{1\leq
j<k\leq N}\epsilon(x_{j}-x_{k})\det_{N}(e^{i x_{m}(\lambda_{n}-q)})
\eeq
where the factor $N!$ comes from the overlap with the basic state
and denominator comes from the normalization.


{\bf The result} for the overlap is given by
\beq
_{C}\langle\psi^{(N)}_{0}(q)|\Psi_{TG}\rangle&=& A_{\lambda}^{(0)}\\
&=&\frac{N!2^{N}\prod_{1\leq j<k\leq
N}(\lambda_{j}-\lambda_{k})}{L^{N}\prod_{i=1}^{N}(\lambda_{i}-q)\prod_{1\leq
i<j\leq N}(\lambda_{i}+\lambda_{j}-2q)}\times \left\{
  \begin{array}{ll}
    (-i)^{\frac{N}{2}}(\cos(\frac{Lq}{2}))^{\frac{N}{2}}(\sin(\frac{Lq}{2}))^{\frac{N}{2}}, & \hbox{N\quad\mbox{even};} \\
    (-i)^{\frac{N-1}{2}}(\cos(\frac{Lq}{2}))^{\frac{N-1}{2}}(-\sin(\frac{Lq}{2}))^{\frac{N+1}{2}}, & \hbox{N\quad\mbox{odd}}
  \end{array}
\right.\nonumber\\
&\times&(-1)^{\sum_{i=1}^{N}n_{i}}\exp(\frac{iL}{2}\sum_{j=1}^{N}\lambda_{j})\exp(-iNL\frac{q}{2})
\label{eq.overlap.delta-p}
\eeq
This formula can be simplified further --- see Sec.~\ref{sec.numeric.delta-p} below.

Straightforward generalization is a construction of a coherent state
in $k$-space. One can take a superposition of
$|\Psi^{(N)}(q)\rangle$ states with different $N$'s for real or
complex $\alpha$:
\beq
|\alpha(q)\rangle =
e^{-\frac{1}{2}|\alpha|^{2}}\sum_{N=0}^{\infty}\frac{\alpha^{N}}{\sqrt{N!}}|\Psi^{(N)}(q)\rangle
\eeq
We will not consider this state here.

\subsubsection{ The overlap with the Gaussian pulse}

In the case of a pulse prepared at time $t=0$ we consider a state
\beq
|\psi_{0}^{(N)}\rangle_{G}=\int \prod_{i=1}^{N} dx_{i}
f(x_{1},\ldots
x_{N})\Psi^{\dag}(x_{1})\ldots\Psi^{\dag}(x_{N})|0\rangle
\eeq
and choose a function $f(x_{1}...x_{N})$ in the form of a product of
gaussians,
\beq
f(x_{1},\ldots
x_{N})=\prod_{i=1}^{N}\exp[-x_{i}^{2}/2\sigma^{2}_{i}]
\eeq
For simplicity, we take $\sigma_{i}=\sigma, \forall i$. The overlap
with the TG state is given by
\beq
\langle\Psi_{TG}|\psi_{0}^{(N)}\rangle_{G} =
\frac{1}{N!}(\frac{\pi}{2})^{N/2}\sigma^{N}
\exp(-\frac{\sigma^2}{2}\sum_{i}\lambda_{i}^{2})\prod_{i}[\mathrm{Erf}(\frac{L-i\lambda_{i}\sigma^{2}}{\sqrt{2}\sigma})+
i\mathrm{Erfi}(\frac{\lambda_{i}\sigma}{\sqrt{2}})]
\label{eq.gauss.overlap}
\eeq
For all practical purposes the Erf function can be replaced by the
gaussian.

\subsection{Tonks-Girardeau limit}
From our understanding of the equilibrium structure factor of the 1D
Bose gas \cite{caux06} we expect that the allowed phase space of the
weakly interacting gas is limited to the vicinity of the dispersion
curve of Bogoliubov quasiparticles. On the other hand we expect that
in the opposite case of the TG limit the allowed phase space should
be considerably larger. Hence non-equilibrium dynamics should be
more involved in this case as well. Here we thus focus on this
interesting limit in more details. Certain simplifications occur in
the equations for the matrix elements, which are described below.

\subsubsection{Reduction of the form factors}

To compute density-density function in the TG limit
\beq
\langle\psi^{(N)}_{0}|\rho(x_{1},t_{1})\rho(x_{2},t_{2})|\psi^{(N)}_{0}\rangle=\frac{1}{L^{2N}N!}\sum_{\lambda,\mu,\nu}F_{N}^{\rho*}(x_{1},t_{1},\lambda,\mu)
F_{N}^{\rho}(x_{2},t_{2},\mu,\nu)A_{\lambda}A_{\nu}
\label{eq.rho_rho.F}
\eeq
we have to know the following form-factor:
\beq
F_{N}^{\rho}(x,t,\{\lambda\},\{\mu\})=e^{it(E_{\lambda}-E_{\mu})}\frac{N}{N!L^{2N}}\int_{0}^{L}dx_{1}\ldots
dx_{N}\sum_{{\cal P,Q}}(-1)^{[{\cal P}]+[{\cal
Q}]}\exp[-ix(\lambda_{{\cal P}}-\mu_{{\cal
Q}})]\\
\exp[-i\sum_{a=1}^{N-1}x_{a}(\lambda_{{\cal P}_{a}}-\mu_{{\cal
Q}_{a}})].
\eeq
A limiting procedure for the form factors in this case has been
developed earlier \cite{CIKT},\cite{IKiS},\cite{CIT}. We use here
the fact that the TG limit can be obtained as a double scaling limit
of the XX chain.  Using
\beq
\int_{0}^{L} dx e^{-ix(\lambda_{a}-\mu_{b})}=\left\{
                              \begin{array}{ll}
                                L, & \lambda_{a}= \mu_{b} \\
                                0, & \lambda_{a}\neq \mu_{b}
                              \end{array}
                            \right.
\eeq
where we use the boundary conditions
$e^{iL\lambda_{a}}=e^{iL\mu_{b}}=(-1)^{N+1}$. Therefore
\beq
F_{N}^{\rho}(x,t,\{\lambda\},\{\mu\})=\left\{
                                        \begin{array}{ll}
                                          L^{N}\rho_{0}, & \{\lambda\}=\{\mu\} \\
                                          L^{N-1}e^{-ix(\lambda-\mu)}, &
\lambda_{1}=\mu_{1},\ldots \lambda_{N-1}=\mu_{N-1},\lambda\neq \mu
                                        \end{array}
                                      \right.=
L^{N}(\rho_{0}\delta_{\lambda\mu}+\frac{e^{-ix(\lambda-\mu)}}{L}(1-\delta_{\lambda\mu}))
\label{eq.F.infty}
\eeq
and 0 otherwise.

Note that this form-factor is antisymmetric function with respect to
any interchange of momenta for both sets $\{\lambda\}$ and
$\{\mu\}$. So when they are not ordered in the formula above, we
have to assign a factor $(-1)^{[P]_{\lambda}+[P]_{\mu}}$, where
$[P]$ is a permutation index.

\subsubsection{Delta function subtraction}
When we calculate density-density correlator~(\ref{eq.rho_rho.F}),
we need to subtract a contribution of the particle with itself. This
contribution comes from commutation relations for Bose operators:
\begin{equation}
\langle\lambda|\Psi^\dag(x_1)\Psi^\dag(x_2)\Psi(x_2)\Psi(x_1)|\nu\rangle
=
\langle\lambda|\Psi^\dag(x_1)\Psi(x_1)\Psi^\dag(x_2)\Psi(x_2)|\nu\rangle
- \langle\lambda|\Psi^\dag(x_1)\Psi(x_1)|\nu\rangle \delta(x_1-x_2)
\end{equation}
Considering the TG limit and using~(\ref{eq.F.infty}) we can
calculate the corresponding $\delta$-function contribution
\begin{equation}
- \langle\lambda|\Psi^\dag(x_1)\Psi(x_1)|\nu\rangle \delta(x_1-x_2).
\end{equation}

For the ground state in the continuous limit this contribution
is given by
\begin{equation}
-\frac{1}{2\pi}\int_{-\infty}^{+\infty} dq e^{i q^2 t - i q x},
\end{equation}
where $x\equiv x_{1}-x_{2}$ and $t\equiv t_{2}-t_{1}$.
In general case~(\ref{eq.rho_rho.F})
delta function yields an additional term
\begin{equation}
- \left(\rho_0 + \frac{1}{L}\sum_{a,b:
\lambda_{a}\neq\nu_{b},\forall b} (-1)^{[P]_\lambda +
[P]_\nu}  A_\lambda A_\nu e^{i x_1 (\lambda_a - \nu_b) - i t_1
(E_\lambda - E_\nu)}\right) \frac{1}{L} \sum_q e^{i x q - i t E_q}.
\end{equation}

Now we can combine all components together to calculate the
density-density correlation function in the TG limit. For the ground
state we have
\begin{equation}
g^{(2)}(\Delta x,\Delta
t)=\rho^{2}+\frac{1}{4\pi^{2}}\int_{|q_{1}|>\pi
\rho}\int_{|q_{2}|<\pi \rho}e^{i
t(q_{1}^{2}-q_{2}^{2})}\cos[(q_{1}-q_{2})x]
-\frac{1}{2\pi}\int_{-\infty}^{+\infty} dq e^{i q^2 t - i q x}.
\label{eq.corr.gs}
\end{equation}
In a general nonequilibrium case correlation function can be written
as
\begin{eqnarray}
\langle\psi^{(N)}_{0}(q)|\rho(x_{1},t_{1})\rho(x_{2},t_{2})|\psi^{(N)}_{0}(q)\rangle
=\rho_{0}^{2}+\frac{1}{L^{2}}\sum_{a,b:
\lambda_{a}\neq\mu_{b},\forall b}
|A_{\lambda}|^{2}e^{ix(\lambda_{a}-\mu_{b})+it(E_{\mu}-E_{\lambda})}\nonumber\\
+\frac{1}{L^{2}}\sum_{a,b,c:
\lambda_{a}\neq\mu_{b\neq\nu_{c}},\forall b,\forall c}
(-1)^{[P]_{\lambda}+[P]_{\nu}+2[P]_{\mu}}A_{\lambda}
A_{\nu}e^{ix_{1}(\lambda_{a}-\mu_{b})+ix_{2}
(\mu_{b}-\nu_{c})+it_{1}(E_{\lambda}-E_{\mu})+it_{2}(E_{\mu}-E_{\nu})}\nonumber\\
+\rho_0 \frac{1}{L} \sum_{a,b:
\lambda_{a}\neq\mu_{b},\forall b} (-1)^{[P]_\lambda +
[P]_\mu}  A_\lambda A_\mu (e^{i x_1 (\lambda_a - \mu_b) - i t_1
(E_\lambda - E_\mu)}+ e^{i x_2 (\lambda_a - \mu_b) - i t_2
(E_\lambda - E_\mu)})\nonumber\\
- \left(\rho_0 + \frac{1}{L}\sum_{a,b:
\lambda_{a}\neq\mu_{b},\forall b} (-1)^{[P]_\lambda +
[P]_\mu}  A_\lambda A_\mu e^{i x_1 (\lambda_a - \mu_b) - i t_1
(E_\lambda - E_\mu)}\right) \frac{1}{L} \sum_q e^{i x q - i t E_q},
\end{eqnarray}
where the first term represents $\delta-\delta$ contribution, the
second one corresponds to $\lambda=\nu$ (diagonal part), third and
fourth terms come from the nondiagonal ($\lambda\neq\mu\neq\nu$)
part, and the last term is a delta contribution discussed above.
Here everywhere $\lambda$ differs from $\mu$ by one filling number
only, and $\mu$ differs from $\nu$ by one filling number as well.

Here, the $\delta$-$\delta$ part as well as diagonal parts are
symmetric with respect to interchange of momenta since these are the
products of two antisymmetric functions. On the other hand in
off-diagonal terms we inserted the sign factors in order to ensure
that proper symmetry is preserved: since each $F$ is antisymmetric,
the \textit{nondiagonal} part will be antisymmetric as well provided
that functions $A_{\lambda}$ and $A_{\nu}$ are symmetric.

The second term (diagonal part) can be simplified further using the
following relation
\beq
\sum_{\lambda_{a}\neq
\mu_{1}\ldots\mu_{N}}f(\lambda_{a})=\sum_{a=1}^{N}f(\lambda_{a})-\sum_{b=1}^{N}f(\mu_{b})
\eeq
for arbitrary function $f$. Then
\beq
\frac{1}{L^{2}}\sum_{a,b: \lambda_{a}\neq\mu_{b},\forall b}
|A_{\lambda}|^{2}e^{ix(\lambda_{a}-\mu_{b})+it(E_{\mu}-E_{\lambda})}=-\frac{1}{L^{2}}|\sum_{a=1}^{N}A_{\lambda}e^{ix\lambda_{a}-itE_{\lambda}}|^{2}\\
+\frac{1}{L^{2}}\sum_{a=1}^{N}\sum_{b=1}^{N}|A_{\lambda}|^{2}e^{ix\lambda_{a}-itE_{\lambda}}e^{-ix\mu_{b}+itE_{\mu}}
\eeq
This identity can be also used to simplify the nondiagonal part.


\subsubsection{Deviation from the TG limit}

The main difficulty of our approach is computing overlaps with the
initial state in a compact, analytical form for arbitrary
interaction strength. At present we do not have a complete solution
of this problem for an arbitrary state. For particular types of
states the problem can be attacked using one of the formalisms of
Section II. Thus, in particular, if the initial state can be
expressed in second-quantized notations via creation-annihilation
operators acting in momentum space, then the second-quantized
version of the intertwining operator can be used. If the initial
state has a simple form in coordinate space, the coordinate version
of the intertwining operator might be useful. One form of this
operator in real space, mostly useful for the $1/c$ expansion around
TG limit, has the form of differential operator acting on the TG
wavefunction \cite{gaudin},\cite{JPGB}. It relates the TG
wavefunction (which is almost fermionic) with the finite-$c$ Bethe
state
\beq
|BA\rangle_{c}=\frac{1}{\sqrt{N!}}\prod_{1\leq i<j\leq
N}[1+\frac{1}{c}(\frac{\partial}{\partial
x_{j}}-\frac{\partial}{\partial
x_{i}})]|\det(e^{i\lambda_{n}x_{n}})|.
\eeq
Manipulations using intertwining operators will be discussed in more
details in the future in connection to concrete physical problems.
Here we note that expanding the intertwining operator up to order
$1/c$ one observes\footnote{We note that
$\int_{0}^{\infty}\int_{0}^{x_{1}}\cdots\int_{0}^{x_{N-1}}\prod_{p=1}^{N}e^{-iqx_{p}}\sum_{n=1}^{N}n\frac{\partial}{\partial
x_{n}}\det(e^{ik_{n}x_{m}}) dx_{N}\cdots
dx_{1}\stackrel{q\rightarrow 0}{=}Ni\frac{\prod_{1\leq i<j\leq
N}(k_{i}-k_{j})\sum_{n=1}^{N}k_{n}}{\prod_{n=1}^{N}k_{n}\prod_{1\leq
i<j\leq N}(k_{i}+k_{j})}$ which is equal then to
$iN\sum_{i=1}^{N}k_{i}A_{\lambda}^{(0)}$, where
$A_{\lambda}^{0}=\int_{0}^{\infty}\int_{0}^{x_{1}}\cdots\int_{0}^{x_{N-1}}\prod_{p=1}^{N}e^{-iqx_{p}}\det(e^{ik_{n}x_{m}})
dx_{N}\cdots dx_{1}\stackrel{q\rightarrow 0}{=}\frac{\prod_{1\leq
i<j\leq N}(k_{i}-k_{j})}{\prod_{n=1}^{N}k_{n}\prod_{1\leq i<j\leq
N}(k_{i}+k_{j})}$. } that a correction to the overlap coefficients
up to the order $1/c$ (neglecting the boundary effects) for the
condensate initial state have the following form
\beq
\tilde{A}_{\lambda}^{(0)}=A^{(0)}_{\lambda}(1+\frac{i(N-1)}{c}P^{(N)}_{\lambda})+o(1/c^{2})
\eeq
Note also that deviations from the TG limit includes other
ingredients. One of them is an expansion of rapidities in powers of
$1/c$,
\beq
\lambda_{j}=(2j-N-1)\frac{\pi}{L}\left[1-2\frac{\rho}{c}+4(\frac{\rho}{c})^{2}-8(\frac{\rho}{c})^{3}+
\frac{4\pi^{2}}{3}N(2j^{2}+(N+1)(N-2j))(\frac{1}{cL})^{3}\right].
\eeq
Using this expression one can get a systematic expansion of the
form-factors. We do not need these $1/c$ corrections since our
procedure uses numerical expressions of the form-factors.

We finally note that a computation of the overlap of the arbitrary
BA state $\langle 0|C(\lambda_{1})\ldots C(\lambda_{n})$ with the
initial state of a special form
$\Psi(x_{1})\ldots\Psi(x_{n})|0\rangle$ can be considered as a
particular case of evaluation of the multiple point form factor if
we regard the pseudo vacuum state as the BA state. The problem of
computation of such a multi point formfactors can be solved using
the formalism of multi-site generalization of the NLS problem. This
problem was addressed in \cite{IKR}. Another way to go beyond the TG
limit is to use different formalisms described in Sec. II.



\section{Numerical treatment}

\subsection{Introduction}

In this part we present results of numerical computations of
correlation functions for various non-equilibrium initial states and
their time evolution. In order to do that, we needed to sum
expression~(\ref{eq.rho_rho}) over all initial $\lambda$,
intermediate $\mu$, and final $\nu$ states.

Although we are interested in thermodynamic limit
$L\rightarrow\infty$, $N\rightarrow\infty$, numerically one can
compute correlation functions for finite size systems only. In order
to infer thermodynamic limit from our finite size results we can set
the size of the problem ($N$ and $L$) to be sufficiently large so
that numerically computed correlation functions would be
sufficiently close to the ones for an infinite system. One of the
criteria for such sufficiency could be comparison of the numeric
results to analytic results for the system for which thermodynamic
solution for correlation functions is exactly known. For example
that we already know the exact expression for a correlation function
in the ground state~(\ref{eq.corr.gs}). So we can use it to check
the accuracy of our numeric procedures.

Let's estimate the cost of a brute force approach towards
summation~(\ref{eq.rho_rho}) for a reasonably large system size.
Because we are interested in
non-equilibrium states, states in $\lambda$, $\mu$, and $\nu$ are
filled way above the ``Fermi momentum''. We can estimate the number
of terms in the sum~(\ref{eq.rho_rho}) by taking $N\sim 100$ and
momentum cutoff
$\sim10N$. Then the number of terms in the sum $\sim \binom{10N}{N}^3\sim
10^{420}$. Obviously, the sum can't be taken in a straightforward
fashion.


Now we would like to look for possible simplifications of the
problem. The first hint comes from~\cite{caux06}, where authors
claim that when one considers summation of the form
\begin{equation}
\sum_\mu |\langle GS|F|\mu\rangle|^2,
\end{equation}
where GS is a ground state, then the sum has a very limited number
of major contributions, most of which are one particle. Note that
this statement is not directly applicable in our case because we are
not working with a ground state and also because we have a double
summation, which might have different major contributions due to larger phase
space, or not to have such major contributions at all.

Second simplification might come from the fact that we are
interested in a large interaction constant $c$ case. For our
purposes we can consider the case of large $c$. As we found in the
previous section, relevant quantities become much simpler. The most
prominent feature of the TG limit is that the only non-zero matrix
elements $\langle\lambda|F|\mu\rangle$ come from zero and one
particle processes~(\ref{eq.F.infty}). It means that in this limit
for a given $\lambda$ we need to do a two particle summation
in~(\ref{eq.rho_rho.F}) with one particle transition
$\lambda\rightarrow\mu$ and one particle transition
$\mu\rightarrow\nu$. This observation also tells us that for large,
but finite $c$, processes with small number of transitions should
dominate.

Finally, structure of overlap factors $A_\lambda$ for different
initial states might provide some clues on how to simplify our task.
In particular for the initial states we are going to investigate,
overlap factors can be simplified analytically so that we do not
need to perform the summation~(\ref{eq.chi.dres})  over all
permutations.

\subsection{Implementation}

Now we would like to discuss ideas behind concrete implementation of
our calculations. Almost everywhere we consider the limit of large
$c$, though we can do numeric calculations for arbitrary $c$ at a
cost of substantially higher cpu time.

For a ground state we do several large-$c$ calculations to check
what values of $c$ can be treated as ``infinite'' and to investigate
multiparticle contributions. For large, but finite $c$, it is
possible to further simplify
expressions~(\ref{Gaudin})-(\ref{eq.Vjpm}) by doing $1/c$ expansion,
as explained in the previous Section. We implemented both direct and
simplified versions of matrix elements and overlap calculations and
did not find any visible discrepancy in the results, though
performance improvement was substantial.

After we sum up the expression
\begin{equation}
\sum_{\lambda,\mu,\nu}A_\lambda A_\nu \langle\lambda|F|\mu\rangle
\langle\mu|F|\nu\rangle
\label{eq.sum}
\end{equation}
over intermediate states $\mu$ and final states $\nu$ for a given
initial state $\lambda$, we are left with a problem of summation
over $\lambda$. But the phase space is huge, therefore we can't use
a direct summation. We use Monte-Carlo summation
instead~\cite{walsh04}.

Namely, we sample set of states $\{\lambda'\}$ from the set of all
possible initial states $\{\lambda\}$ with probability
of state $\lambda'$ to be selected being $P_\lambda$.
Then we can replace summation in~(\ref{eq.sum}) over $\lambda$ with
summation over $\lambda'$ to get
\begin{equation}
I^\mathrm{MC}=\frac{ \sum_{\lambda',\mu,\nu}A_{\lambda'} A_\nu
\langle\lambda'|F|\mu\rangle \langle\mu|F|\nu\rangle/P_{\lambda'}}
{\sum_{\lambda',\mu,\nu}A_{\lambda'} A_\nu /P_{\lambda'}}.
\label{eq.sum.mc}
\end{equation}
As we increase the size of sample $\{\lambda'\}$, the value of this
sum converges to the true value of~(\ref{eq.sum}) for any
probability distribution $P_\lambda$. Depending on the pick of the
probability distribution function convergence time can be very
different. For instance if we take a uniform distribution over the
entire set $\{\lambda\}$, than we can see from
Eq.~(\ref{eq.gauss.overlap}) that weights $A_\lambda$ can be very
different. As a result, many contributions to the~(\ref{eq.sum.mc})
will be negligibly small. On the other hand if we are picking
$\lambda$ such that $|A_\lambda|$ is around its maximum, all the
contribution to the sum will be substantial.

Indeed, it is known~\cite{walsh04}, that the sum optimally converges
to the true value if the distribution function is proportional to
the expression we sum. In our case it would (roughly speaking) mean
$P_\lambda\propto A_\lambda\sum_\nu A_\nu$, where $\nu$ differs from
$\lambda$ at most by two particle process. As a reasonable
approximation, we use $P_\lambda\propto A_\lambda^2$.

The next question is how to sample such states $\lambda'$.
We cannot sample states from the distribution directly, but we can use Gibbs
sampling~\cite{geman84} to do that approximately.
Note that any overlap factor can be expressed as a function of state
$\lambda$ and parameters of the problem. $\lambda$ itself is a function
of a vector of quantum numbers $\vec{I}$~(\ref{BAeq}). We can fix all the components
of $\vec{I}$ except one ($I_j$). By varying this component and resolving the
equation~(\ref{BAeq}) relative to $\lambda$, we can find conditional probabilities
$P(I_j|I_1\ldots I_{j-1}I_{j+1}\ldots I_n)\propto A^2_{\lambda|I_j}$, where
$\lambda|I_j$ is a solution for $\lambda$ given quantum number $I_j$ (and the
rest of quantum numbers, which stay the same).
This way we can sample $I_j$ from the $P_\lambda$ distribution given
all other quantum numbers $I_{i\neq j}$ are fixed.
Then we iterate the procedure with index $j$ running from 1 to $N$ several times.
Using this algorithm we obtain a set of statistically independent
states pooled with probability $\propto A_\lambda^2$~\cite{geman84}.

In theory one can do Monte-Carlo summation not only over $\lambda$ states,
but over $\mu$ and $\nu$ as well. Nevertheless, we found such approach
impractical because of a much slower convergence stemmed from very
poor cancellation of various harmonics.

\subsection{Results}

\subsubsection{Ground state}

First we would like to focus on correlation functions for the ground
state. We have several reasons for that. First, exact analytical
solution for such correlation functions is known~(\ref{eq.corr.gs}) , hence we can
at least partially verify the validity of our approach. Second, we do
not have to sum over $\lambda$ and $\nu$, so computations are very
fast and precise.
Therefore we can use this case to investigate the effects of our
approximations --- what finite values of interaction constant $c$
can be considered as infinite, how one particle approximation
affects results, and what are the finite size effects of our
calculations.

\begin{figure}
\includegraphics[angle=270,width=12.5cm]{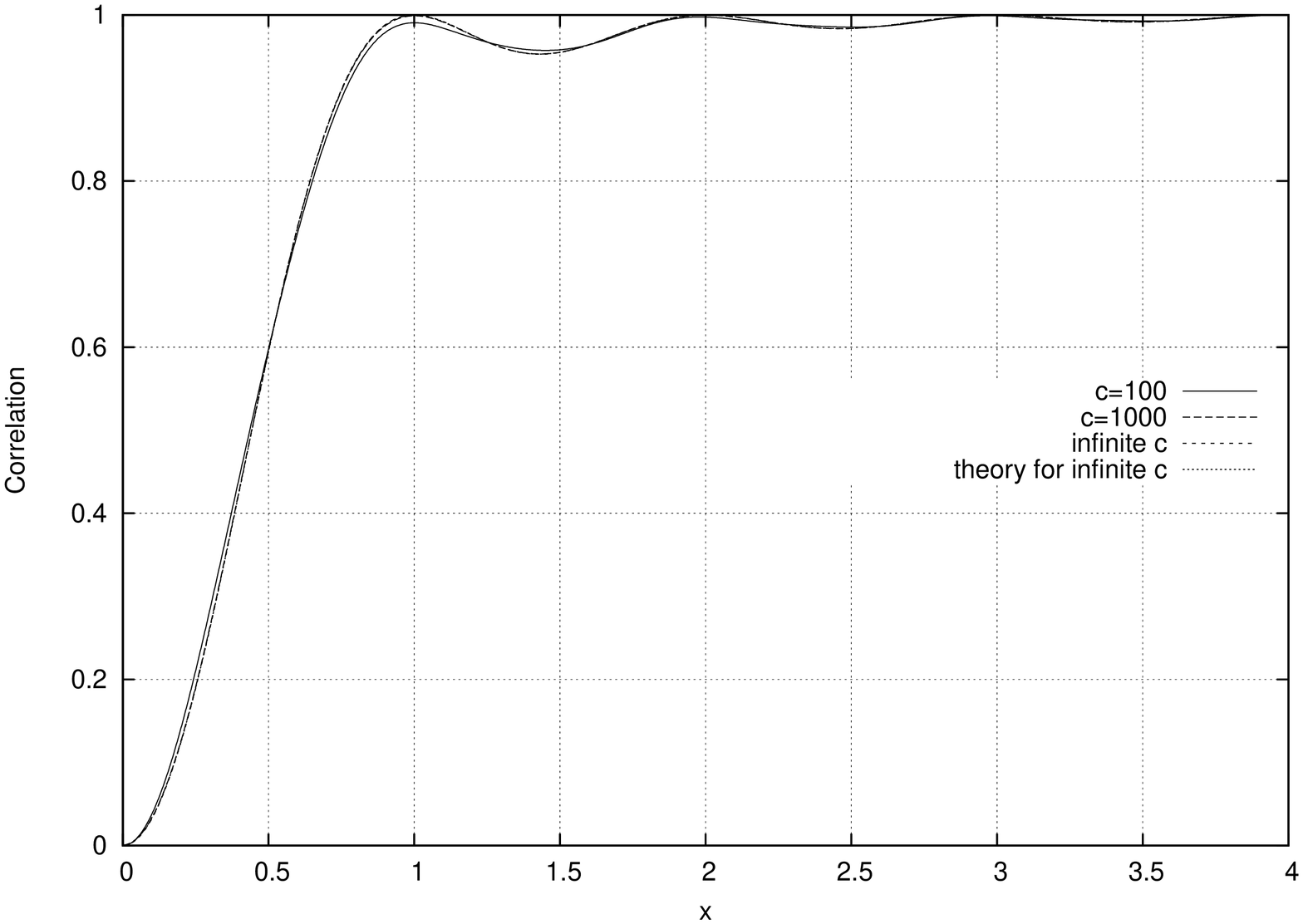}\\
\includegraphics[angle=270,width=12.5cm]{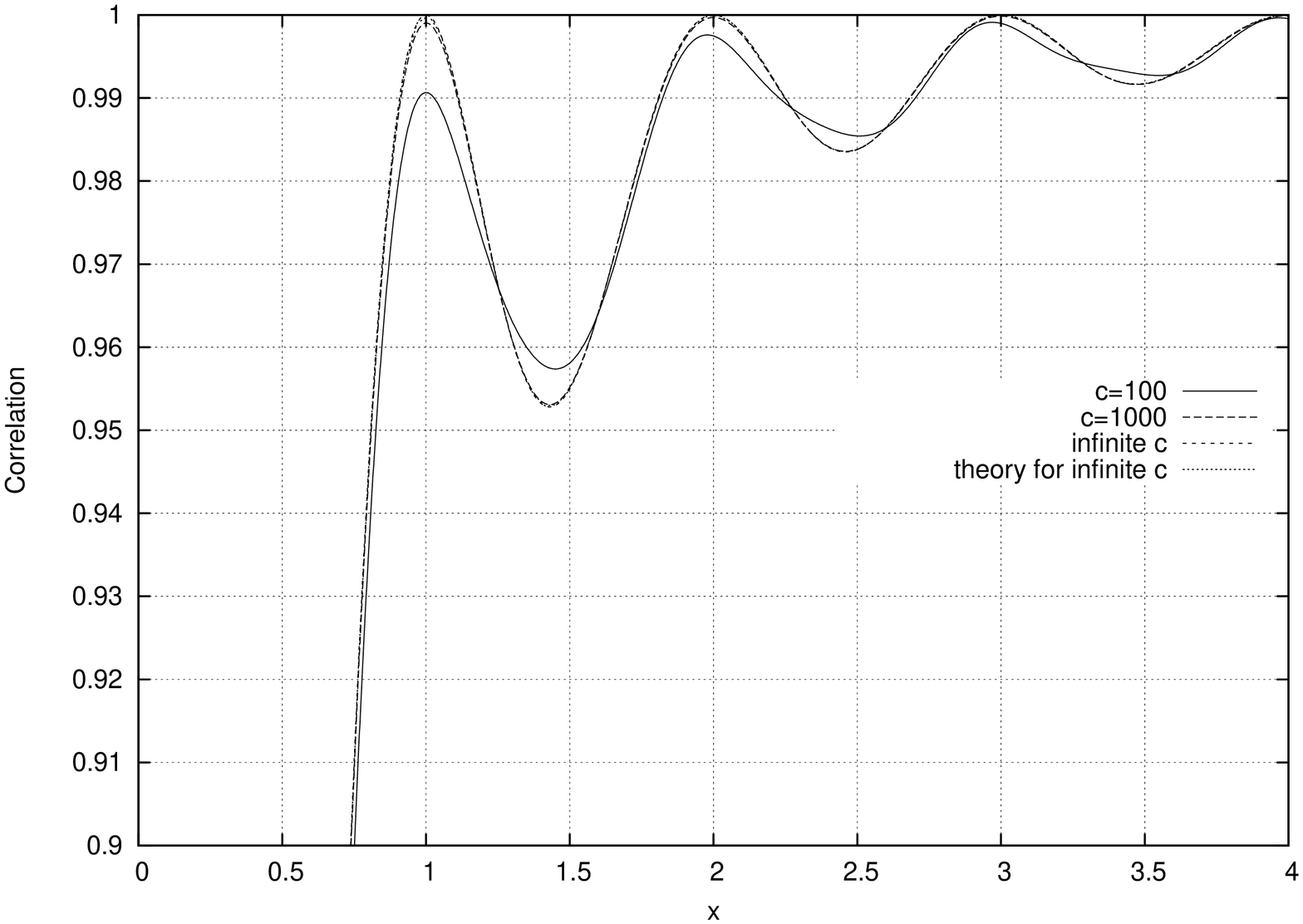}
\caption{Upper panel: calculated and theoretical spatial correlation
function for a ground state for $c=100$, $c=1000$, and
$c\rightarrow\infty$. Lower panel: zoom of the upper panel.}
\label{fig.ground.spatial}
\end{figure}

First we looked at spatial correlator, calculated at $L=51$, $N=51$.
For the ground state it is translationally invariant both in space
and time. On Figs.~\ref{fig.ground.spatial} we show a theoretical
correlator along with ones  calculated at $c=100$, $c=1000$, and
$c=\infty$ for one particle processes only. We see that for $c=1000$
the correlator is almost indistinguishable from the theoretical
$c\rightarrow\infty$ limit, though $c=100$ slightly deviates from
that limit. Hence all our further $c\rightarrow\infty$ results apply
to $c\gtrsim 1000$ case as well. Also note that theoretical
correlation function perfectly overlaps with one calculated for
$c=\infty$, hence our choice of system size can be considered as a
thermodynamic limit.

\begin{figure}
\includegraphics[angle=270,width=12.5cm]{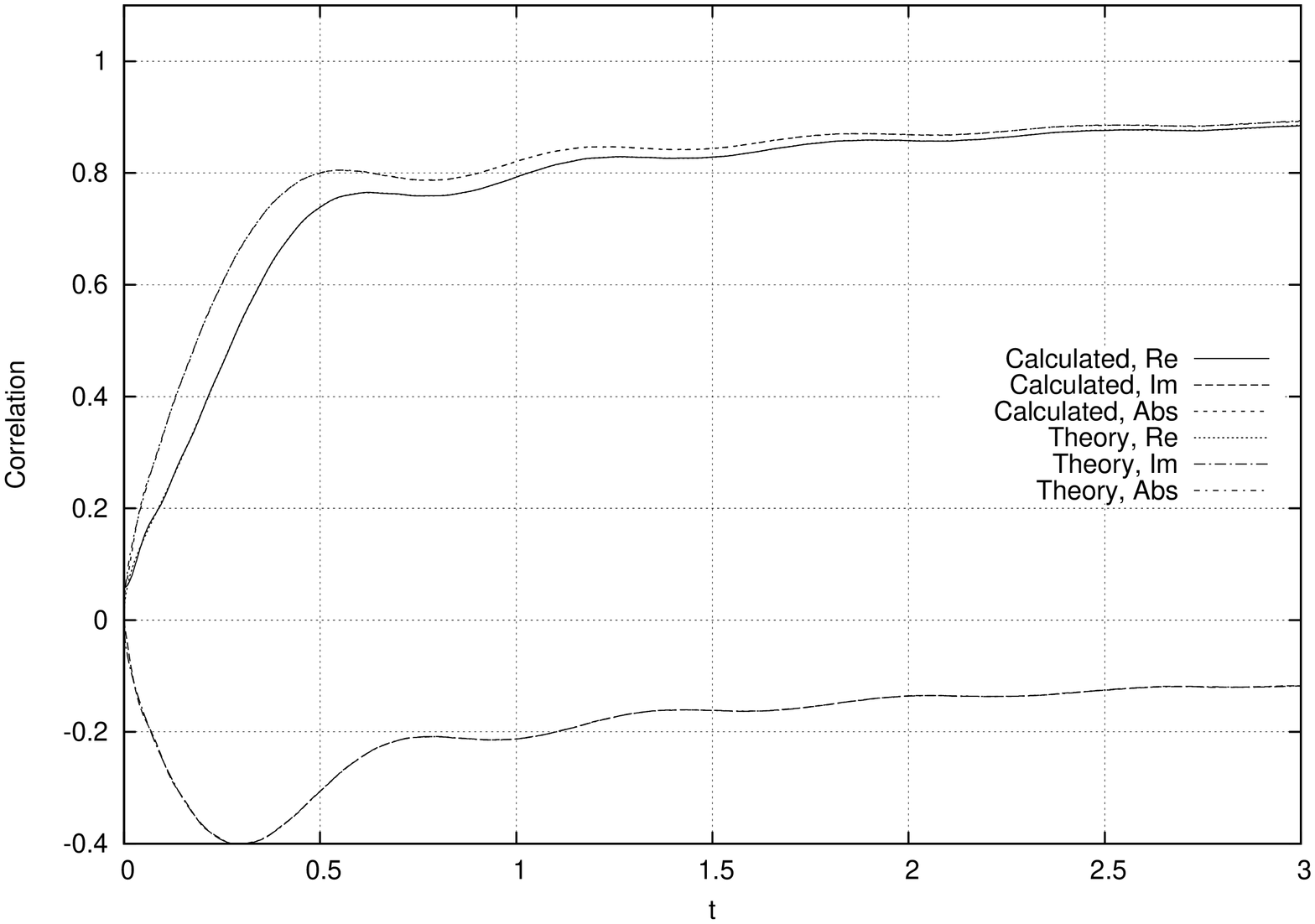}
\caption{Theoretical and calculated temporal correlator for a ground
state in the $c\rightarrow\infty$ limit. Real, imaginary parts, and
absolute values are shown. Only three curves are visible because
theoretical and calculated plots perfectly overlap.}
\label{fig.ground.temporal}
\end{figure}
As a separate run we calculated the same correlators for the same
finite values of $c$ while taking into account all two-particle
contributions and major three-particle contributions in addition to
the original one-particle ones. We found that these contributions
are negligible. We conclude that it is safe to ignore them in the
large $c$ limit.

On Fig.~\ref{fig.ground.temporal} we show a temporal correlation
function for a ground state for $c\rightarrow\infty$ as well as a
theoretical value given by~(\ref{eq.corr.gs}). We see that they
perfectly overlap except for the tiny region around x=0 because of
momentum cutoff in our numeric calculations.

\subsubsection{Delta function in momentum space}
\label{sec.numeric.delta-p}

We consider the case when before interactions are switched on all
particles are initially in the same state with a given momentum $p$.
Such state can be prepared experimentally from the condensate at
rest using a Bragg pulse.

Before we proceed with calculations, we would like to take a closer
look at overlap factors $A_\lambda$ for this
state~(\ref{eq.overlap.delta-p}). We are starting with the
equation for the projection of our state onto eigenfunctions (TG) of
the Hamiltonian $\langle\Psi^{(N)}(q)|\Psi_{TG}\rangle$. Ignoring
coefficients independent of the TG state and non-singular at $q=0$,
we have for odd~\footnote{These calculations can be easliy repeated
for even $N$ resulting in the same conclusion.}
$N$
\begin{equation}
\langle\Psi^{(N)}(q)|\Psi_{TG}\rangle \propto \frac{\prod_{1 \leq j
< k \leq N}(\lambda_j-\lambda_k)}
{\prod_{i=1}^N(\lambda_i-q)\prod_{1 \leq j < k \leq
N}(\lambda_j+\lambda_k-2q)}
\left(\sin\left(\frac{Lq}{2}\right)\right)^\frac{N+1}{2}.
\label{eq.overlap}
\end{equation}
Note that our system is discretely translationally invariant
in $k$ space: we can shift all $\lambda_i$ and $q$ by the same
constant $\frac{2\pi}{L}n$ for arbitrary $n$, change
quantum numbers $I_j$ by $n$, and all the
equations (\ref{BAeq})-(\ref{eq.overlap.delta-p}) will hold.

Neglecting finite size effects (resulting in discretization), we can
choose $n$ such that $q=\frac{2\pi}{L}n$. Then we shift all
$\lambda_i$ by $q$, and consider the overlap~(\ref{eq.overlap})
factor in the $q\rightarrow 0$ limit.

\begin{figure}
\includegraphics[angle=270,width=12.5cm]{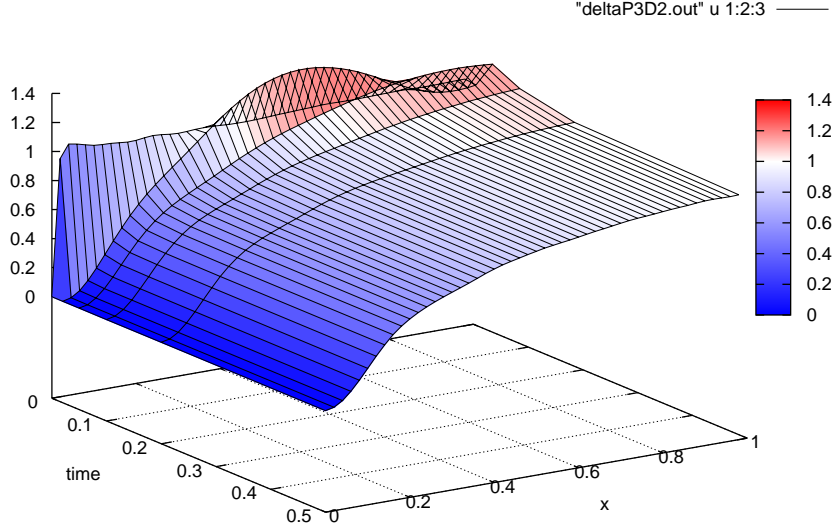}
\caption{Three-dimensional image (coordinate, time and intensity) of
the correlation function for the initial state corresponding to the
delta peak in momentum space.}
\label{deltaP3D}
\end{figure}

Sinus in the expression~(\ref{eq.overlap}) results in a
$\frac{N+1}{2}$ order zero in the numerator in $q\rightarrow\infty$
limit.
Zeroes of the denominator at $q=0$ are given by
one of $\lambda_i=0$ in the first product,
and all the pairs $(i,j), i<j$ such that
$\lambda_i = -\lambda_j$ in the second product.
Maximum order of zero we can achieve in
the denominator is when we have one zero $\lambda_i$, and for each of
the rest of lambdas we have one exactly opposite to it. For such
configuration zero's order of the denominator is also
$\frac{N+1}{2}$. In this case they cancel resulting in a non-zero
contribution. Any other configurations will result in the lower
order of zero of the denominator, hence zero overlap.
We conclude that all the states with non-zero overlap are
symmetric. One can easily repeat this calculation for even
$N$ and show that states in this case are symmetric as well.

We can draw an immediate conclusion about the structure of the
sum~(\ref{eq.sum.mc}) we use to calculate correlation function. Remember
that for an infinite $c$ $\langle\lambda|F|\mu\rangle$ has non-zero elements
only if states $\lambda$ and $\mu$ differ by at most one quantum number~(\ref{eq.F.infty}).
Therefore in the summation~(\ref{eq.sum.mc})
$\lambda$ is symmetric, $\mu$ has one quantum number changed
relative to $\lambda$, and $\nu$, being symmetric, has one quantum
number changed relative to $\mu$. But it is possible only if either
$|\lambda\rangle=|\nu\rangle$, or if the changed quantum number in
$|\lambda\rangle\rightarrow|\mu\rangle$ is symmetric to the
corresponding change for $|\mu\rangle\rightarrow|\nu\rangle$. I.e.
if first we change $I_i\rightarrow I'_i$, than later we should
change either $I'_i\rightarrow I_i$ or $-I_i=I_j\rightarrow
I'_j=-I'_i$. As a result for a given $|\lambda\rangle$
a selection of $|\mu\rangle$ allows only
two possible states for $|\nu\rangle$, strongly reducing the phase
space we need to consider.

Using the symmetry we can also simplify the
expression for overlap factors~(\ref{eq.overlap.delta-p}).
Because in the sum~(\ref{eq.sum.mc}) normalization of $A_\lambda$
does not matter, we drop all the constant terms and can
consider~(\ref{eq.overlap}) instead.
Let's denote $n=(N-1)/2$ ($N$ is
odd), and indexes run from $-n$ to $n$. Then we can rewrite the
numerator as
\begin{equation}
\prod_{0<j<k\leq n}(\lambda_j-\lambda_k)^2(\lambda_j+\lambda_k)^2
\prod_{0<j\leq n}-2\lambda_j^3
\end{equation}
and the denominator as
\begin{equation}
(-q)^\frac{N+1}{2}2^\frac{N-1}{2}\prod_{0<j<k\leq
n}(\lambda_j-\lambda_k)^2(\lambda_j+\lambda_k)^2 \prod_{0<j\leq
n}\lambda_j^4,
\end{equation}
so that~(\ref{eq.overlap.delta-p}) simplifies to
\begin{equation}
A_\lambda\propto\frac{1}{\prod_{\lambda_j>0}\lambda_j}.
\label{eq.overlap.simpl}
\end{equation}
In case of interest one can easily restore the constant coefficient and also
prove that this expression is valid for even $N$ as well.

\begin{figure}
\includegraphics[angle=270,width=8.5cm]{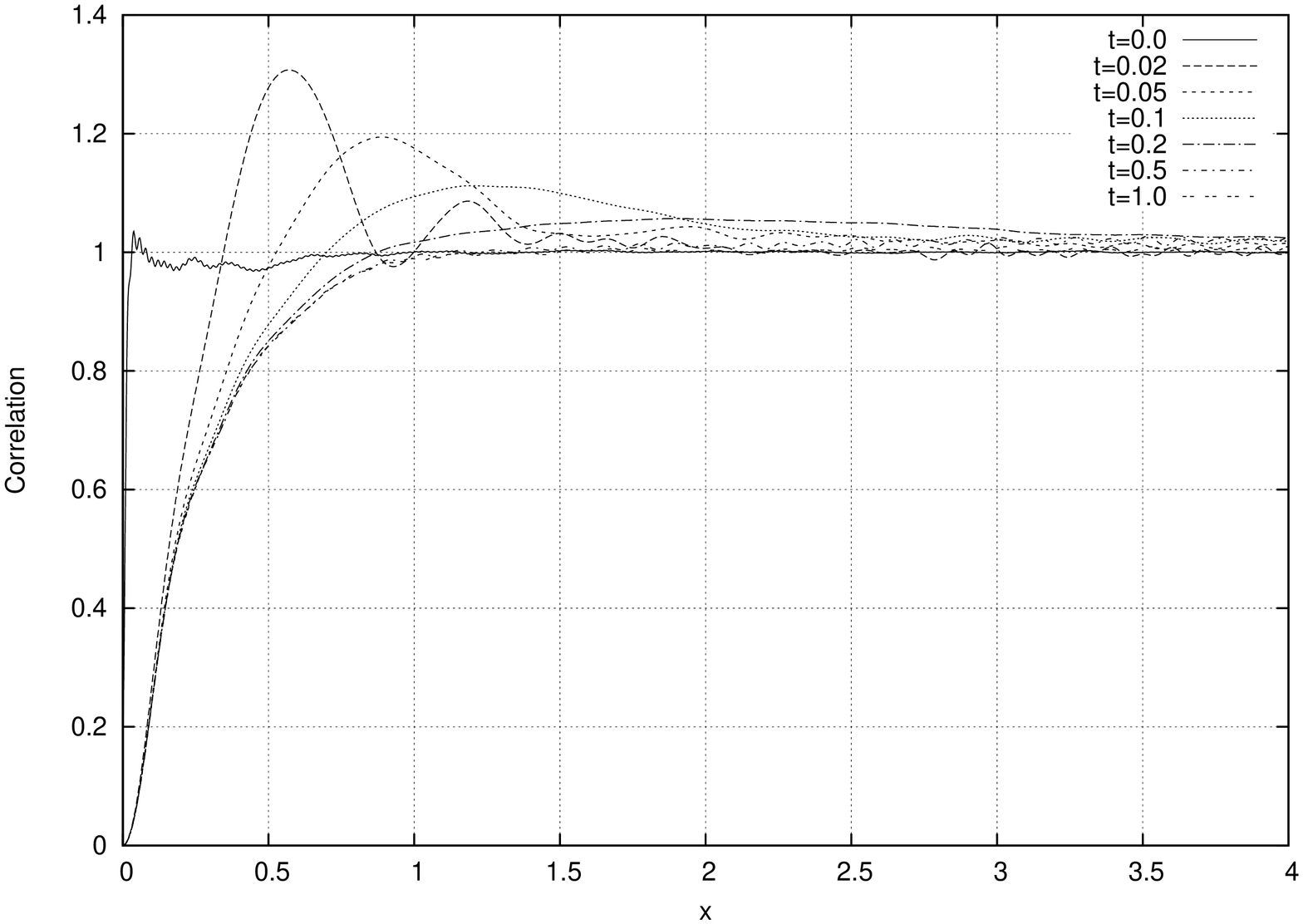}\includegraphics[angle=270,width=8.5cm]{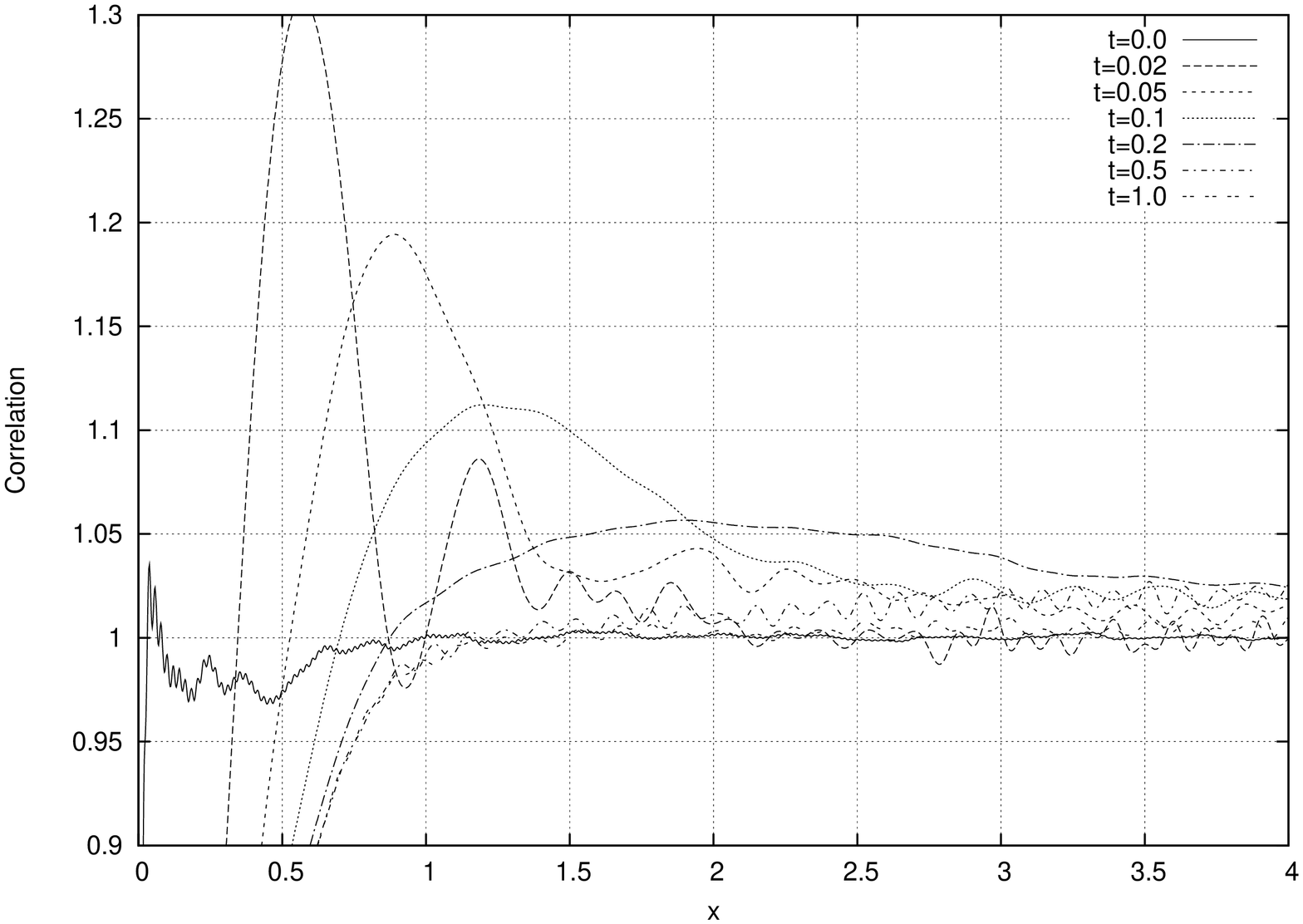}
\caption{Left: spatial correlator for the initial state
corresponding to the delta function in momentum space for different
$t_0$. Right: zoom of the left panel.}
\label{fig.pdelta.spatial}
\end{figure}

Using the conclusions drawn above about relative structure of
$\lambda$, $\mu$, and $\nu$ states, and using
expression~(\ref{eq.overlap.simpl}) for overlap factors, we perform
a Monte Carlo summation over states for $N=101$, $L=101$. On
Fig.~\ref{fig.pdelta.spatial}
we show a spatial correlation $\langle
\rho(x_0,t_0)\rho(x_0+x,t_0)\rangle$ as a function of $x$ for
$x_0=0$ and different values of $t_0$. Basically the plot represents
time evolution of $x$ correlator starting from the moment we
momentarily switch on a very strong interaction. We can see that at
the moment $t_0=0$ the entire correlation function is not distorted
except for $x=0$. This is because when we immediately switch on the
interaction, particles stay where they are. They just don't have any
time to move anywhere, hence the original non-interacting correlation
function, which is flat, is preserved (small distortions
around $x=0$ are artifacts of our calculations; if to take into account
their errors, they are indistinguishable from 1). But as the time goes by,
repelling interaction generates a wave, which transfers the matter
away from the particle sitting at $x=0$. On the figure we observe
the evolution of this wave. When the time becomes sufficiently large
($t\gtrsim 1$), the correlation function stabilizes and does not
evolve anymore.

%
\begin{figure}
\includegraphics[angle=270,width=12.5cm]{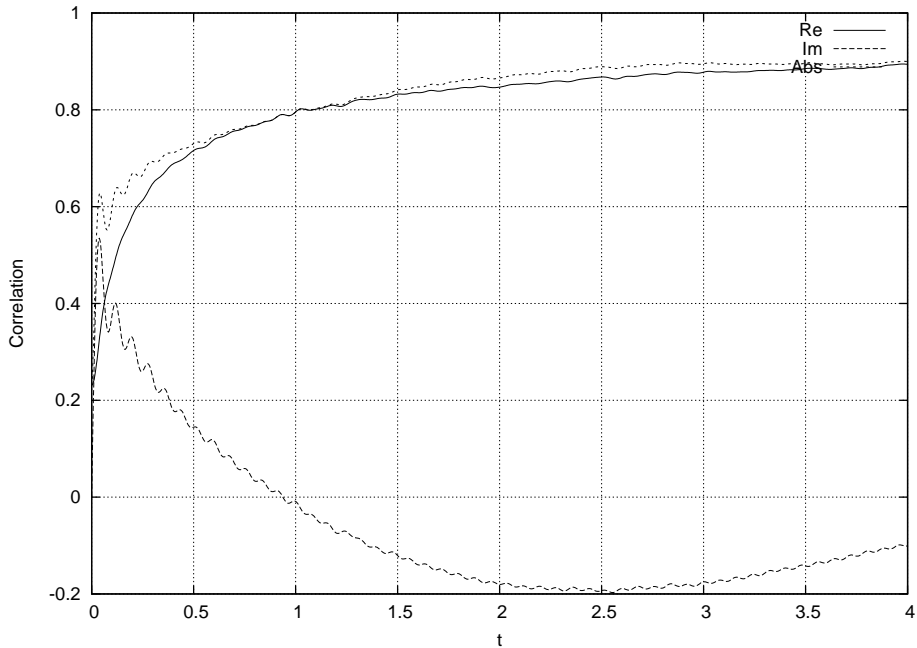}
\caption{Real and imaginary parts of a temporal correlator for the
initial state corresponding to the delta function in momentum
space.}
\label{fig.pdelta.temporal}
\end{figure}

We also calculate a temporal correlator $\langle
\rho(x_0,t_0)\rho(x_0,t_0+t)\rangle$.
Figure~\ref{fig.pdelta.temporal} shows real and imaginary parts of
the correlator. Because the state is translationally invariant,
correlator does not depend on $x_0$. Interestingly, it also almost
does not depend on $t_0$. All the plots for different $t_0$
overlapped, and the difference was invisible.
On this plot we also observe two ``effects'', which are artifacts
of our calculation scheme because of introduced momentum cutoffs:
oscillations in the imaginary part and real part
not going exactly to zero at $t=0$.

\subsubsection{Gaussian pulse}

Now we proceed with the initial state --- a Gaussian pulse in
space, and hence in momentum space. Overlap factors for such state are given
by the expression~(\ref{eq.gauss.overlap}). We neglect finite size effects
by considering the limit $L\gg\sigma$, for which
expression~(\ref{eq.gauss.overlap}), up to a constant factor, can be written
as
\begin{equation}
A_\lambda \propto \exp\left(-\frac{\sigma^2}{2}\sum_i \lambda_i^2 \right).
\label{eq.gauss.overlap.simpl}
\end{equation}

In this case we do not have a nice symmetry we had in the previous section,
so we have to sum over all one particle $\mu$ states and two
particle $\nu$ states using expression~(\ref{eq.sum.mc}) . Summation over all two
particle processes substantially increases the computational time,
so in order to partially compensate for it we used smaller values
for problem parameters $L=21$ and $N=21$. For a packet width parameter we use value
$\sigma=15/(2\pi)$.

\begin{figure}
\includegraphics[angle=270,width=\columnwidth]{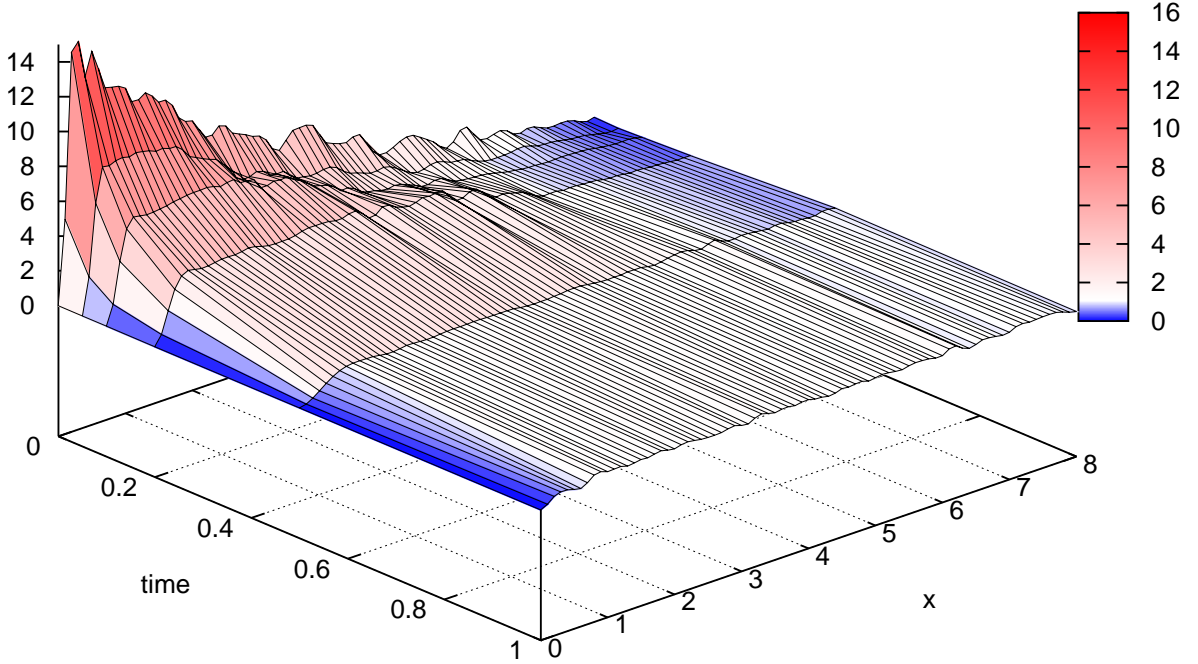}
\caption{Evolution of the correlation function for the initial state
corresponding to the Gaussian in the real space.}
\label{gauss3D}
\end{figure}
\begin{figure}
\includegraphics[angle=270,width=12.5cm]{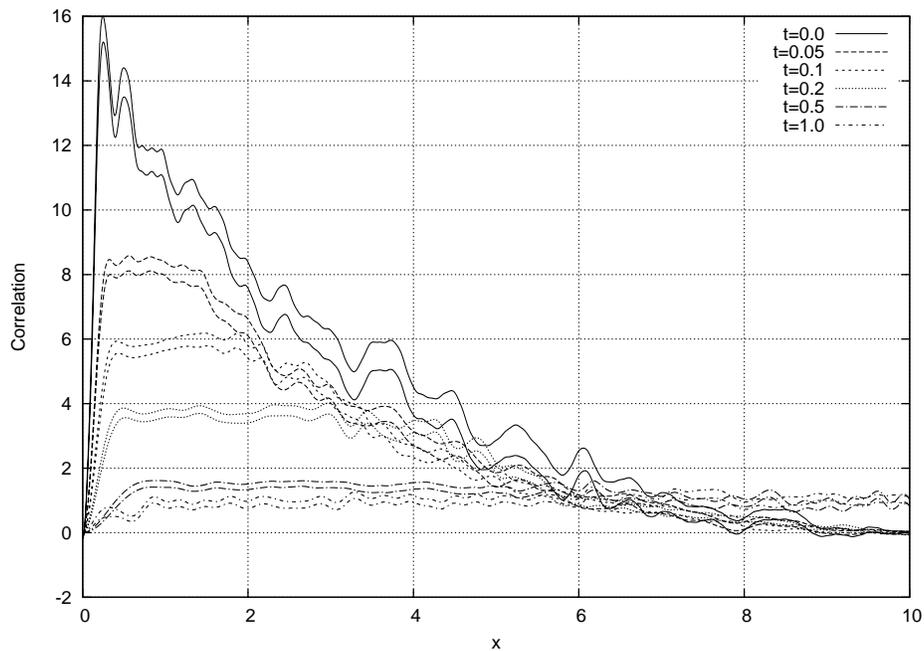}
\caption{Spatial correlator for different $t_0$ for a Gaussian
initial state and its error. Two lines for each value of $t_0$ lie
one standard deviation above and below relative to the average
value.}
\label{fig.gauss.spatial}
\end{figure}
On Fig.~\ref{fig.gauss.spatial} we present time evolution of spatial
correlator $\langle\rho(x_0,t_0)\rho(x_0+x,t_0)\rangle$ for $x_0=0$
and different fixed times $t_0$. We also show errors of our
calculations to distinguish true correlation function features from
noise.

We would like to emphasize several features visible on these plots.
First, for $t_0=0$, as expected, we see a dip at $x=0$.
Interestingly, the width of this dip is finite and
is not determined by the
momentum cutoff. From computational point of view we can
explain it by the fact that the overlap factor~(\ref{eq.gauss.overlap.simpl})
for Gaussian
state strongly inhibits high harmonics, therefore we do not have
momenta high enough to make this dip infinitesimally narrow.
This is a finite size effect because of a small number of particles.
As parameters $L$, $N$, and $\sigma$ increase proportionally, width
of the dip goes to zero.
 Second, oscillations of the correlator at $t_0=0$ around the dip $x_0=0$
are not determined by the cutoff and much bigger than the error of
our calculations. This is again the effect of a finite size system.
Third, as we've already seen in previous section,
for times $t_0\gtrsim 1$ the
correlation function stabilizes. But the process of stabilization is
quite remarkable. We do not have a widening of a Gaussian peak.
Instead it looks like the peak has the same width, but its height
decreases with time. Probably, qualitative explanation lies in the
fact that high density regions have high interaction energies, hence
in these regions we have fast particles which quickly leave the
region. I.e. we have a ballistic scenario, which differs from the
diffusion scenario for which we would expect a widening of a
Gaussian peak with time.

Unfortunately, due to some peculiarities of computations we had to
conduct, we were unable to produce reliable results for a temporal
correlator. Results, though reproducing an overall shape of the
correlator, were very noisy because of beats attributed to non-exact
cancellation of various harmonics. We do not provide a plot for
these results.

\section{Discussion and Summary}
In this paper we addressed a problem of non-equilibrium time
evolution of one dimensional Bose gas with contact interaction from
the general perspective of dynamics of integrable systems. Several
approaches have been proposed recently for analyzing time evolution
of integrable many-body models, including Quantum Inverse Scattering
method, the formalism of Intertwining Operator, and Extended
Conformal Symmetry. After critically reviewing these approaches we
concentrated on conceptually the simplest method, based on using
Bethe ansatz to decompose initial states into precise eigenstates of
the interacting Hamiltonian using the intertwining operator and then
using form factor expansion to calculate time evolution of
correlation functions. The main difficulty of this method is that it
requires summation over a large number of intermediate states. In
this paper we focused on the regime of strong repulsive interactions
between bosons and developed an efficient numerical procedure for
performing summation over intermediate states. We analyzed two types
of initial states: all particles having zero momentum and all
particles in a gaussian wavepacket in real space. In both cases we
find a non-trivial time evolution of the second order coherence
$g_2(x_1,x_2,t)=\langle\rho(x_1,t),\rho(x_2,t)\rangle$, which
reflects the intrinsic dichotomy between initial states and the
Hamiltonian. Initial condensate states exhibit bunching at short
distances, whereas strong repulsion in the Hamiltonian introduces
strong antibunching.

For the initial state which has all particles in a state with zero
momentum, results for $g_2$ are summarized in Fig.~\ref{deltaP3D}.
Antibunching at shortest distances is present at all times
reflecting repulsion between particles. A more striking feature is
the appearance of bunching and oscillations in $g_2(x_{1}-x_{2},t)$
at intermediate time and length-scales. At transient times $g_2$ has
Friedel like oscillations as a function of $x_{1}-x_{2}$, which
disappear at longer times. Crystallization of the TG gas in the
process of non-equilibrium time evolution was also discussed in Ref
\cite{Chang}. At longer times the form of $g_2$ is qualitatively
similar to what one finds for the equilibrium Lieb Liniger model
with intermediate interactions at high temperatures \cite{SCDVRK}.
Fig.~\ref{deltaP3D} also provides some support to the idea of light
cone formation discussed in Ref.~\cite{CCcone}. Some of these
features are similar to quench dynamics in other integrable systems
\cite{BRDMBG,FCC}.

For the initial state which has all particles in a gaussian
wavepacket in real space, time evolution of $g_2$ is shown in
Figs.~\ref{gauss3D},\ref{fig.gauss.spatial}. As in the previous case
we observe antibunching at the shortest distances. For intermediate
time scales it is followed by a pronounced peak in $g_2$ reflecting
dynamic bunching. This system would be naturally characterized by
the time dependent {\it short range correlation length}, which
increases with time, reflecting expansion of the system in real
space. Oscillations in $g_2$ which we observe in this case are small
and are most likely related to the finite number of particles used
in the analysis. Hence transient states of this system would be more
natural characterized as a liquid rather than a crystal.

Predictions made in our paper for the behavior of $g_2$ should be
possible to test in experiments with ultracold atoms and photons in
strongly non-linear medium.

\section{Acknowledgement}
We would like to thank J.-S. Caux and M. Zvonarev for useful
discussions. V.G. and E. D. are supported by AFOSR, DARPA, MURI, NSF
DMR-0705472, Harvard-MIT CUA, and Swiss National Science Foundation.

\section{Appendix A: Inverse Scattering Transform and Algebraic
Bethe Ansatz - Basic concepts}

The nonlinear Schr\"{o}dinger Hamiltonian represents the simplest
example of a system solvable by the algebraic Bethe ansatz
\cite{Sklyanin1,Faddeev}. For review and many details see
\cite{BIK}. Here we overview the basic construction with connection
to the inverse scattering transform in order to fix notations and
for the sake of self-consistency.

The Zakharov-Shabat method starts from transforming the field
$\Psi(x,t=0)$ of nonlinear Hamiltonian (\ref{Hc}) at time $t=0$ into
a set of "scattering data" given by the following linear problem
\beq
i\frac{\partial}{\partial x}\Phi(x,\xi)=Q(x,\xi)\Phi(x,\xi)
\eeq
where
\beq
Q(x,\xi)=\left(
           \begin{array}{cc}
             -\frac{\xi}{2} & -\sqrt{c}\Psi(x) \\
             \sqrt{c}\Psi^{\dag}(x) & \frac{\xi}{2} \\
           \end{array}
         \right)
\eeq
The solution $\Phi(x,\xi)$ of this equation is defined by the
condition $|\Psi(x)|\rightarrow 0$ as $x\rightarrow\pm\infty$ and by
the properties of Jost solutions,
\beq
\left(
  \begin{array}{c}
    1 \\
    0 \\
  \end{array}
\right)e^{i\xi
x/2}&\mathop{\longleftarrow}_{x\rightarrow-\infty}&\left(
                                                     \begin{array}{c}
                                                       \phi_{1}(x,\xi) \\
                                                       \phi_{2}(x,\xi) \\
                                                     \end{array}
                                                   \right)\mathop{\longrightarrow}_{x\rightarrow\infty}\left(
                                                                                            \begin{array}{c}
                                                                                              A(\xi)e^{i\xi x/2} \\
                                                                                              B(\xi)e^{-i\xi x/2} \\
                                                                                            \end{array}
                                                                                          \right),\\
\left(
  \begin{array}{c}
    -B^{\dag}(\xi)e^{i\xi x/2} \\
    A(\xi)e^{i\xi x/2} \\
  \end{array}
\right)&\mathop{\longleftarrow}_{x\rightarrow-\infty}&\left(
                                                     \begin{array}{c}
                                                       \chi_{1}(x,\xi) \\
                                                       \chi_{2}(x,\xi) \\
                                                     \end{array}
                                                   \right)\mathop{\longrightarrow}_{x\rightarrow\infty}\left(
                                                                                            \begin{array}{c}
                                                                                              0e^{i\xi x/2} \\
                                                                                              1e^{-i\xi x/2} \\
                                                                                            \end{array}
                                                                                          \right)e^{-i\xi
                                                                                          x/2}
                                                                                          \eeq

Rewriting the linear equation above with the boundary conditions as
an integral equation gives the Gelfand-Levitan-Marchenko equation
\beq
\xi_{1}(x,\xi)e^{ix\xi/2}=-\sqrt{c}\int_{-\infty}^{\infty}dy\theta(y-x)e^{i\xi
y}\Psi^{\dag}(y)\chi_{2}(y,\xi)e^{-iy\xi/2},\\
\xi_{2}(x,\xi)e^{-ix\xi/2}=1+i\sqrt{c}\int_{-\infty}^{\infty}dy\theta(y-x)e^{-i\xi
y}\chi_{1}(y,\xi)\Psi(y)e^{iy\xi/2}
\eeq
which can be solved iteratively in powers of $\sqrt{c}$
\beq
\chi_{1}(x,\xi)e^{ix\xi/2}&=&-i\sqrt{c}\sum_{n=0}^{\infty}c^{n}\prod_{i=0}^{n}\int
dx_{i}\prod_{j=1}^{n}dy_{j}\theta(x_{0}>y_{1}>\cdots
y_{n}>x_{n}>x)e^{i\xi(\sum_{i=0}^{n}x_{i}-\sum_{j=1}^{n}y_{j})}\\
&\times&\Psi^{\dag}(x_{0})\cdots\Psi^{\dag}(x_{n})\Psi(y_{n})\cdots\Psi(y_{1}),\\
\chi_{2}(x,\xi)e^{-ix\xi/2}&=&1+\sum_{n=0}^{\infty}c^{n}\prod_{i=0}^{n-1}\int
dx_{i}\prod_{j=1}^{n}dy_{j}\theta(x_{0}>y_{1}>\cdots
x_{n-1}>y_{n}>x)e^{i\xi(\sum_{i=0}^{n-1}x_{i}-\sum_{j=1}^{n}y_{j})}\\
&\times&\Psi^{\dag}(x_{0})\cdots\Psi^{\dag}(x_{n-1})\Psi(y_{n})\cdots\Psi(y_{1})
\eeq
where
$\theta(x_{1}>x_{2}>\cdots>x_{n})=\theta(x_{1}-x_{2})\theta(x_{2}-x_{3})\cdots\theta(x_{n-1}-x_{n})$.
This solution can be written as a solution for the scattering data,
$A(\xi)$ and $B(\xi)$. Finally, defining the reflection operator as
\beq
R(\xi)=i[A(\xi)]^{-1}B(\xi)
\eeq
one obtains the expansion (\ref{glmexp}). Operators $R(\xi)$ satisfy
the Zamolodchikov-Faddeev algebra (\ref{Rbasis}). In this formalism,
the Bethe ansatz states are constructed by application of
Zamolodchikov-Faddeev operators to the pseudovacuum state
$|0\rangle$,
\beq\label{BAstateR}
|\psi_{N}\rangle_{BA}=R^{\dag}(\xi_{1})\ldots
R^{\dag}(\xi_{N})|0\rangle
\eeq
These states are complete for the case of repulsion. Using the
expansion (\ref{glmexp}) one can show \cite{CTW} that the {\it
special} ordered initial state of the form
\beq\label{ord_state}
|\psi_{N}\rangle_{ord}=\theta(x_{1}>x_{2}>\ldots>x_{N})|\Psi^{\dag}(x_{1})\Psi^{\dag}(x_{2})\ldots\Psi^{\dag}(x_{N})|0\rangle
\eeq
is equal to the following state
\beq
R^{\dag}(x_{1})R^{\dag}(x_{2})\ldots R^{\dag}(x_{N})|0\rangle
\eeq
where
\beq
R(x)=\int\frac{d\xi}{2\pi}e^{ix\xi}R(\xi).
\eeq
We note that the evolution of this state can be obtained explicitly,
as discussed in Section II. The case of attraction is discussed in
Appendix C.

The transfer matrix is a central ingredient of the construction of
the inverse scattering transform on the finite interval on the
lattice (one performs a space discretization procedure for the LL
model first). It can be represented as a matrix
\beq
T(\xi)=\left(
  \begin{array}{cc}
    A(\xi) & B(\xi) \\
    C(\xi) & D(\xi) \\
  \end{array}
\right)
\eeq
The transfer matrix is constructed as a product of monodromy
matrices $L_{j}(\xi)$ for each site $j$,
$T(\xi)=\prod_{j}L_{j}(\xi)$, which in the case of NS problem has
the form of matrix $Q$ which is one of matrix forming the Lax pair.
Trace of the transfer matrix commute for different $\xi$ and is a
generator of integrals of motion. The integrability condition is
expressed by the special property
\beq
R(\xi-\xi')(T(\xi)\otimes T(\xi'))=(T(\xi')\otimes
T(\xi))R(\xi-\xi')
\eeq
where the matrix $R(\xi,\xi')$ is a solution of the Yang-Baxter
equation, which for the case of NS problem belongs to the class of
6-vertex model
\beq
R_{12}(\xi)R_{13}(\xi+\mu)R_{23}(\mu)=R_{23}(\mu)R_{13}(\xi+\mu)R_{12}(\xi)\\
R_{ab}=\beta I_{ab}+\alpha P_{ab},\qquad \alpha
=\frac{\xi_{a}-\xi_{b}}{\xi_{a}-\xi_{b}-ic},\qquad\beta=\frac{-ic}{\xi_{a}-\xi_{b}-ic}
\eeq
where $I_{ab}$ and $P_{ab}$ are identity and permutation operators
acting on the tensor product of two single-particle spaces indexed
by $a$ and $b$. In the finite size lattice formalism the diagonal
entries of the matrix (more precisely its trace
$\tau(\xi)=TrT(\xi)=A(\xi)+D(\xi)$) generate a conserved quantities
of the model whereas the off-diagonal elements $B(\xi)$ and $C(\xi)$
act as creation and annihilation operators of pseudoparticles. The
$n$-particle Bethe ansatz eigenvectors are constructed using these
off-diagonal elements of the transfer matrix as
\beq
|\Psi_{n}(\{\lambda_{n}\})\rangle=B(\lambda_{1})B(\lambda_{2})\ldots
B(\lambda_{n})|0\rangle
\eeq
whereas the bra-vectors are constructed as
\beq
\langle\Psi_{n}(\{\lambda_{n}\})|=\langle
0|C(\lambda_{1})C(\lambda_{2})\ldots C(\lambda_{n})
\eeq
The algebraic Bethe ansatz deals with diagonalization of the trace
$\tau(\xi)$ of the transfer matrix, thus solving the eigenvalue
equation $\tau(\lambda)|BA\rangle
=\Lambda(\lambda,\{\lambda_{a}\})|BA\rangle$. This can be done using
the commutation relations between operators
$A(\lambda),B(\lambda),C(\lambda),D(\lambda)$ which can be deduced
from the Yang-Baxter algebra. This procedure leads to expression for
the
$\Lambda(\lambda,\{\lambda_{a}\})=\alpha(\lambda)\prod_{a=1}^{n}f(\lambda-\lambda_{a})+\beta(\lambda)\prod_{a=1}^{n}f(\lambda_{a}-\lambda)$
which can be further related to the energy and to the consistency
relation, known as a Bethe ansatz equations for the momenta
$\lambda_{a}$ of the pseudo particles,
\beq
\frac{\alpha(\lambda_{a})}{\beta(\lambda_{a})}=\prod_{b\neq
a}\frac{\lambda_{b}-\lambda_{a}}{\lambda_{a}-\lambda_{b}},
\eeq
where the function $f(\lambda)$ is determined by the elements of the
$R$-matrix. The connection with the continuum version of the NLS
model is established by a limiting procedure of lattice size going
to zero while keeping the density constant on a finite interval.

Due to the work of Sklyanin \cite{Sklyanin2} it is possible to
formulate even more general NS problem which includes the
boundaries. The boundary problems can be divided into two
categories, soliton-preserving and soliton-non-preserving. For
recent progress in solution of the boundary NS model see
\cite{Doicou}.


\end{document}